\newif\ifconfver
\confvertrue        %declaring conference version u

\ifconfver
\documentclass[10pt,twocolumn,twoside]{IEEEtran}
\else
\documentclass[11pt,draftcls,onecolumn]{IEEEtran}
\fi

\usepackage{makecell}
\usepackage{bbding}
\usepackage{color,graphicx}
\usepackage{float}
\usepackage{multirow,amsmath,epsfig,amsfonts,amsbsy,amssymb,psfig,graphics,psfrag,theorem,calc,url,bm,cite}
\usepackage{stfloats,hyperref}
\usepackage{algorithm,algorithmic}
\usepackage{diagbox} 
\usepackage{hhline}
\usepackage{subfigure}
\usepackage{comment}
\usepackage{tikz}
\usepackage{quantikz}
\usepackage{pdfrender}
\newtheorem{theorem}{Theorem}
\usepackage{mathtools}

\makeatletter
\def\multilimits@{\bgroup
	\Let@
	\restore@math@cr
	\default@tag
	\baselineskip\fontdimen10 \scriptfont\tw@
	\advance\baselineskip\fontdimen12 \scriptfont\tw@
	\lineskip\thr@@\fontdimen8 \scriptfont\thr@@
	\lineskiplimit\lineskip
	\vbox\bgroup\ialign\bgroup\hfil$\m@th\scriptstyle{##}$\hfil\crcr}
\def\Sb{_\multilimits@}
\def\endSb{\crcr\egroup\egroup\egroup}
\makeatother

{\begin{list}{}%
		{\setlength{\rightmargin}{0pc}%
			\setlength{\leftmargin}{1pc} }} %
	{\end{list}}

\newlength{\twidth}
\ifconfver
\else
\fi
\definecolor{orange}{RGB}{255,107,0}

\theorembodyfont{\rmfamily}

{\begin{list}{}{
			\settowidth{\labelwidth}{\mbox{\textnormal{#1}}}%
			\setlength{\leftmargin}{\labelwidth+\labelsep}}}%
	{\end{list}}

\newcommand\bA{\ensuremath{{\bm A}}}
\newcommand\bB{\ensuremath{{\bm B}}}

\newcommand\bD{\ensuremath{{\bm D}}}

\newcommand\bI{\ensuremath{{\bm I}}}

\newcommand\bM{\ensuremath{{\bm M}}}
\newcommand\bN{\ensuremath{{\bm N}}}

\newcommand\bS{\ensuremath{{\bm S}}}

\newcommand\bY{\ensuremath{{\bm Y}}}
\newcommand\bZ{\ensuremath{{\bm Z}}}

\newcommand\ba{\ensuremath{{\bm a}}}

\definecolor{orange}{RGB}{255,107,0}

\ifconfver
\author{Chia-Hsiang Lin,~\IEEEmembership{Senior Member,~IEEE}, and Jhao-Ting Lin,~\IEEEmembership{Student Member,~IEEE}}
\title{PRIME: Blind Multispectral Unmixing Using

~~~~Virtual Quantum Prism and Convex Geometry
    \thanks{This study was supported by the Emerging Young Scholar Program (namely, the 2030 Cross-Generation Young Scholars Program) of National Science and Technology Council (NSTC), Taiwan, under Grant NSTC 113-2628-E-006-003.
    We thank the National Center for Theoretical Sciences (NCTS) and the National Center for High-performance Computing (NCHC) for providing the computing resources.}
    \thanks{\textit{(Corresponding author: Chia-Hsiang Lin.)}}
    \thanks{C.-H. Lin is with the Department of Electrical Engineering, and with the Miin Wu School of Computing, National Cheng Kung University, Tainan, Taiwan (R.O.C.) (e-mail: chiahsiang.steven.lin@gmail.com).}
    \thanks{J.-T. Lin is with the Institute of Computer and Communication Engineering, Department of Electrical Engineering, National Cheng Kung University, Tainan, Taiwan (R.O.C.) (e-mail: q38091534@gs.ncku.edu.tw).}
}

\else

\fi

\begin{document}
	
	\bibliographystyle{IEEEtran}
	\maketitle
	\ifconfver \else \vspace{-0.5cm}\fi
	
\begin{abstract}
Multispectral unmixing (MU) is critical due to the inevitable mixed pixel phenomenon caused by the limited spatial resolution of typical multispectral images in remote sensing.
However, MU mathematically corresponds to the underdetermined blind source separation problem, thus highly challenging, preventing researchers from tackling it.
Previous MU works all ignore the underdetermined issue, and merely consider scenarios with more bands than sources.
This work attempts to resolve the underdetermined issue by further conducting the light-splitting task using a network-inspired virtual prism, and as this task is challenging, we achieve so by incorporating the very advanced quantum feature extraction techniques.
We emphasize that the prism is virtual (allowing us to fix the spectral response as a simple deterministic matrix), so the virtual hyperspectral image (HSI) it generates has no need to correspond to some real hyperspectral sensor; 
in other words, it is good enough as long as the virtual HSI satisfies some fundamental properties of light splitting (e.g., non-negativity and continuity).
With the above virtual quantum prism, we know that the virtual HSI is expected to possess some desired simplex structure.
This allows us to adopt the convex geometry to unmix the spectra, followed by downsampling the pure spectra back to the multispectral domain, thereby achieving MU.
Experimental evidence shows great potential of our MU algorithm, termed as prism-inspired multispectral endmember extraction (PRIME).

%---------
\end{abstract}

	\ifconfver \else \vspace{-0.0cm}\fi
	
	\ifconfver \else \vspace{-0.5cm}\fi
	
	%ken does some tricks with line spacing: end here
	\ifconfver \else  \fi

\section{Introduction} \label{sec: intro}

Most existing optical satellites acquire multispectral images (MSIs) over diverse landscapes, while they often suffer from the so-called mixed pixel phenomenon, as typical MSIs in remote sensing have limited spatial resolution.
This means that the spatial region of one pixel could cover a large area, hence containing multiple substances, and thus a pixel could be considered as a mixture of several pure multispectral spectra of the underlying substances.
To facilitate practical applications, we need to develop blind source separation (BSS) technology \cite{HISUN} to unambiguously recover the pure spectra from the mixed multispectral pixels for better classification and identification.
To echo relevant BSS development in the hyperspectral remote sensing area \cite{bioucas2012hyperspectral}, we term this BSS technology as multispectral unmixing (MU).

However, the pioneering MU work \cite{MU1} does not consider the underdetermined scenarios wherein there are more sources than observations; for example, one of the experiments therein involves $P=8$ multispectral bands but only with $N=4$ sources.
To understand that our work aims at solving the MU from a fundamentally different but more practical perspective, let us mathematically describe our problem as to recover the multispectral endmember matrix $\bB\in\mathbb{R}^{P\times N}$ and the abundance matrix $\bS\in\mathbb{R}^{N\times L}$ from the observable MSI $\bZ_m=\bB\bS\in\mathbb{R}^{P\times L}$ in the case of more sources than bands (i.e., $P<N$), where $P$ is the number of multispectral bands, $N$ is the number of sources/substances presented in $\bZ_m$, and $L$ is the number of pixels.
Note that what we are interested in is the more practical underdetermined MU case of $P<N$, because typical multispectral satellites often have less than $5$ spectral bands, while a typical MSI often covers more than $5$ substances.
However, previous MU works all ignore the underdetermined issue, and merely consider scenarios with more bands than sources (i.e., $P>N$).

With a fundamentally different approach, this work attempts to resolve the underdetermined issue by incorporating a virtual prism to further conduct the light-splitting function on the observed MSI.
Inspired by our recent investigation on the spectral super-resolution technique \cite{lin2022fast}, such a virtual light-splitting task is considered technically feasible.
As a side remark, the spectral super-resolution in \cite{lin2022fast} is achieved by blending the advantages of both convex optimization (CO) and deep learning (DE), leading to the so-called CODE learning theory \cite{ADMM-ADAM}.
Let us get back to the required virtual prism.
Since the light-splitting function is challenging, this paper tries to achieve so through the help of the very advanced quantum feature extraction techniques \cite{li2020quantum}.
Specifically, we implement this prism function as a quantum deep network (QUEEN), and, to ensure that our overall MU algorithm is a \textit{blind} one, we have to train the QUEEN in a fully unsupervised manner.
Note that blind mechanism is often preferred in remote sensing \cite{PWTdehazing}.
In very recent remote sensing literature \cite{HyperQUEEN}, QUEEN has been applied to well restore satellite-acquired hyperspectral images (HSIs).
As this technique, referred to as hyperspectral QUEEN (HyperQUEEN), has a very solid mathematical ground (e.g., provable quantum full expressibility), its quantum network architecture will be employed in our MU method \cite{HyperQUEEN}.

To facilitate the above idea, a basic light-splitting function for reducing the complexity could already be sufficient for the MU application.
Specifically, unlike the light-splitting method proposed in \cite{lin2022fast} achieving approximately $10$ times spectral super-resolution, the virtual prism in our MU method is set by default to perform $2$ times spectral super-resolution because usually we have $2P\geq N$ in practice (e.g., a $4$-band MSI covering $6$ material substances).
However, as will be experimentally demonstrated, our MU method can be easily upscaled to perform higher order spectral super-resolution, if necessary (cf. Section \ref{sec: discussion}).
Here, we purposely emphasize that the prism is virtual (allowing us to fix the spectral response as a simple deterministic matrix \cite{COCNMF}), so the prism-generated virtual HSI has no need to correspond to some real hyperspectral sensor.
Therefore, it should be sufficient as long as the virtual HSI satisfies some fundamental properties of light splitting (e.g., non-negativity and spectral continuity), greatly simplifying the design of the prism architecture and thus increasing the technical feasibility of the proposed idea.
With the above virtual quantum prism, the prism-generated virtual HSI is expected to possess some natural properties of HSI data, such as the desired simplex geometry structure \cite{MVSA}.
This not only allows us to utilize the virtual HSI to judiciously formulate the MU problem, but also allows us to solve the problem by adopting convex geometry theory to effectively perform HU on the virtual HSI, followed by spectrally downsampling the extracted pure endmember spectra back to the multispectral domain to complete the MU task.
Experimental evidences demonstrate the feasibility and superiority of our proposed MU algorithm.

The remaining parts of this paper are organized as below.
In Section \ref{sec: related}, we review existing HU methods, highlighting their theoretical foundations and key advancements.
Additionally, given the limited MU literature, we discuss relevant underdetermined BSS techniques to echo MU.
In Section \ref{sec: algo}, we illustrate our innovative prism-based idea for solving the underdetermined MU problem, and show how it induces the cycling data-fitting term in our proposed MU criterion.
We also implement the prism using quantum deep learning, and accordingly develop an algorithm to realize the proposed criterion, where we plug a convex geometry mechanism to achieve efficient computing.
As this is the first MU work that fundamentally considers the underdetermined scenario, we design an experimental protocol to evaluate the MU performance in Section \ref{sec: experiment}, where we demonstrate the superiority of our algorithm with ablation study.
Finally, the conclusions are drawn in Section \ref{sec: conclusion}.

Some standard notations used in this paper are collectively presented hereinafter.
$\mathbb{Z}_{++}$ denotes the sets of positive integers.
$\mathbb{R}^n$ and $\mathbb{R}^{m\times n}$, denote, respectively, real-valued $n$-dimensional vectors and real-valued $m$-by-$n$ matrices.
The set difference is denoted as ``$\setminus$".
$\|\cdot\|_F$ and $\|\cdot\|_1$ denote, respectively, the Frobenius norm and the $\ell_1$-norm.
The symbols $\bm 1_m$ and $\bm I_{m}$ represent, respectively, an all-one column vector of dimension $m$ and an identity matrix of dimension $m$-by-$m$.
The Kronecker product is defined as $\otimes$.
Given a matrix $\bM$, the notation $\bM\geq\bm{0}$ means that $\bM$ is elementwise non-negative; $[\bM]_{(i,:)}$, $[\bM]_{(:,j)}$, and $[\bM]_{(i,j)}$ denote the $i$th row of $\bM$, $j$th column of $\bM$, and the $(i,j)$th entry of $\bM$, respectively.
$\textrm{DIAG}(\bY_1,\dots,\bY_N)$ denotes the block-diagonal matrix with $\bY_n$ being the $n$th diagonal block for $n=1,\dots,N$ \cite{CVXbookCLL2016}.
The convex hull, affine hull \cite{CVXbookCLL2016}, and volume of a given set $\mathcal{S}$ are, respectively, represented as $\textrm{conv}\mathcal{S}$, $\textrm{aff}\mathcal{S}$,  and $\textrm{volume}(\mathcal{S})$.

\section{Related Works} \label{sec: related}
%SUnSAL C-SUnSAL 2010
%SUnSAL-TV 2012
%SUnSPI 2015
%SUnCNN 2022
%SFSU 2024
In this section, we begin by reviewing existing HU methods, quite related to MU.
With the increase in open spectral libraries, sparse regression (SA), a semi-supervised approach, has recently been proposed to address the HU problem, effectively reducing reliance on pure pixels.
Sparse unmixing by variable splitting and augmented Lagrangian (SUnSAL) \cite{SUnSAL} and its constrained variant, C-SUnSAL, are classical methods based on the alternating direction method of multipliers (ADMM).
Building on SUnSAL, SUnSAL-TV \cite{SUnSAL-TV} introduces total variation regularization to impose spatial consistency.
SUnSPI \cite{SUnSPI} incorporates spectral a priori information to promote sparsity in inactive signatures.
Spatial filtering-based sparse unmixing (SFSU) \cite{SFSU} adopts the edge-preserving filtering strategy to address complex noise interference.
With the rapid advancements in artificial intelligence (AI), SUnCNN \cite{SUnCNN} emerged as the first DE technique for sparse unmixing, utilizing a convolution neural network (CNN) architecture combined with a softmax activation layer to enforce the abundance sum-to-one constraint.

While the above approaches effectively enhance abundance estimation, geometry-based methods that operate in a fully blind manner (i.e., performing endmember extraction without much prior knowledge) are often preferred in remote sensing due to their broader applicability.
Under the so-called pure-pixel assumption, vertex component analysis (VCA) \cite{VCA} iteratively identifies the farthest pixel (i.e., one of the pure pixels) from the data centroid, then projects the data onto a direction orthogonal to the subspace spanned by the previously selected pixels.
N-FINDR \cite{N-FINDR} employs Winter's criterion to identify the largest simplex formed by the pure pixels within the data convex hull \cite{CVXbookCLL2016}. 
However, a significant challenge of the pure-pixel assumption lies in the mixed pixel phenomenon, which makes pure pixels sparse or absent, thereby limiting the effectiveness of these algorithms.

In contrast to Winter's criterion, Craig's criterion \cite{Craig1994,HyperCSI,packer2002np} recovers the pure spectra using the vertices of the minimum simplex that contains all the data pixel points.
Non-negative matrix factorization (NMF) \cite{NMFcomplexity} is well-suited for HU due to its strong correlation with the low-rank nature of HSIs.
By incorporating Craig's criterion as a regularization term, minimum volume constrained NMF (MVC-NMF) \cite{MVC-NMF} suppresses noise sensitivity and improves the accuracy of endmember and abundance estimation.
Since Craig's criterion and NMF often involve NP-hard criteria, John ellipsoid criterion \cite{MVIE} has been proposed as an alternative approach for high-performance HU, aiming to mitigate the NP-hardness.
John ellipsoid criterion can be implemented using convex optimization, with convex log-determinant objective function over the positive definite cone, and with the convex second-order cone constraints.
This convex problem can be solved using the first-order optimizer within polynomial time, based on the fast iterative shrinkage-thresholding algorithm (FISTA) \cite{FISTA}.

Unlike HU, which typically involves more observations than sources, the MU problem mathematically corresponds to the underdetermined equation system and is theoretically much more challenging than HU.
This heightened complexity likely prevents researchers from tackling MU, making it quite rarely seen in the literature.
After conducting a detailed survey, we found one probably the most relevant work \cite{MU1}, where some seminal HU algorithms are extended for MU.
Specifically, MU is attempted by the three HU algorithms, including VCA \cite{VCA}, N-FINDR \cite{N-FINDR}, and NMF \cite{NMF-HU}.
The three algorithms are selected partly because they (or their variations) are the most widely used unmixing methods, which are proven to be robust and reliable, and partly because they represent different HU categories showing unique advantages depending on the real scenarios \cite{MU1}.

Though there are some underdetermined BSS methods attempting to solve problems similar to MU, they are developed for speech mixture signals (not applicable for image-type signals), and would make very specific sparsity assumptions on the sources \cite{underdeterminedBSS1,underdeterminedBSS2}.
There are very few BSS methods designed for the underdetermined case.
Most underdetermined BSS methods are designed for acoustic signals (not non-negative).
For example, in \cite{sawada2011underdetermined}, the authors proposed a clustering-based method in the frequency domain that can be applied to a scenario where there are fewer microphones than sources.
%============================
In \cite{ma2010novel}, the authors adopt the singular spectrum analysis to approach underdetermined BSS, where the analysis requires the precondition of stationary and statistically independent sources.
Nevertheless, in remote sensing images, the sources are seldom statistically independent \cite{nascimento2005does}.
%============================

\section{The Proposed MU Algorithm} \label{sec: algo}

\subsection{Criterion Design}

Remotely sensed multispectral or hyperspectral satellite images often suffer from the so-called mixed pixel phenomenon.
HU is thus widely studied to recover pure hyperspectral signatures (endmembers) for effective material identification.
However, MU is seldom seen in the literature because it corresponds to the much more challenging underdetermined case.
Though MU seems impossible, we trickily employ the recently sprouted spectral super-resolution technology \cite{lin2022fast} and quantum technology \cite{HyperQUEEN} to reformulate MU into a virtual HU problem.
We solve the reformulated HU to obtain virtual endmembers, which are then transformed back to the multispectral domain, thereby obtaining the MU solution.
According to the above solution outline, let us introduce some notations below.

Given a typical HSI $\bZ_h\in\mathbb{R}^{M\times L}$, the linear mixing model can be adopted to describe it as $\bZ_h=\bA\bS$ \cite{VCA,2024SISHY}, where $\bA\in\mathbb{R}^{M\times N}$ is the hyperspectral endmember matrix and $\bS\in\mathbb{R}^{N\times L}$ is the abundance matrix.
Here, $M$ is the number of hyperspectral bands, $L$ is the number of pixels, and $N$ is the number of materials presented in $\bZ_h$.
As for the physical meaning, the $i$th column of $\bA$ is the hyperspectral signature of the material $i$, while the $i$th row of $\bS$ corresponds to the abundance map of the material $i$, for $i=1,\dots,N$.
For most benchmark HU methods, $N$ is assumed known a priori, because the model-order selection itself is a challenging issue \cite{lin2017MDL}.
As discussed in Section \ref{sec: intro}, MU is regarded as more challenging than HU.
So, in this work, we also assume that $N$ is known, and focus on the tough MU problem itself.

To describe the associated MSI $\bZ_m$, its relation with $\bZ_h$ can be modeled as $\bZ_m=\bD\bZ_h\in\mathbb{R}^{P\times L}$, where $P$ is the number of multispectral bands, and $\bD\in\mathbb{R}^{P\times M}$ is the spectral response matrix \cite{COCNMF}.
So, we have $\bZ_m=\bD\bA\bS=\bB\bS$, where $\bB\triangleq\bD\bA\in\mathbb{R}^{P\times N}$ is referred to as the multispectral endmember matrix.
Here, the $i$th column of $\bB$ is the multispectral signature of the material $i$, for $i=1,\dots,N$.
The MU problem can now be mathematically described as to recover the two unknown quantities $(\bB,\bS)$ from the MSI $\bZ_m=\bB\bS$ when $P<N$, which is clearly an instance of underdetermined BSS.
We remark that though this sounds like a typical NMF problem, it is actually more challenging as typical NMF often considers the case of $P>N$ \cite{NMF1999}.

As discussed in Section \ref{sec: intro}, we have noticed some works claiming that they are able to solve MU, but their scenarios are actually with $P\geq N$.
Also, though there are some underdetermined BSS techniques for $P<N$, they are not applicable to MU.
Therefore, this work is regarded as the first attempt to explore the possibility of solving MU directly for the underdetermined case of $P<N$.
To this end, an intuitive criterion is to minimize the naive NMF data-fitting term, i.e.,
\begin{equation}
    \|\bZ_m-\bB\bS\|_F=\|\bZ_m-\bD\bA\bS\|_F,
\end{equation}
which is, however, ineffective as it does not resolve the underdetermined issue (still less bands than sources).

\begin{figure}[t]
    \begin{center}
    \resizebox{0.7\linewidth}{!}{\hspace{-0cm}\includegraphics{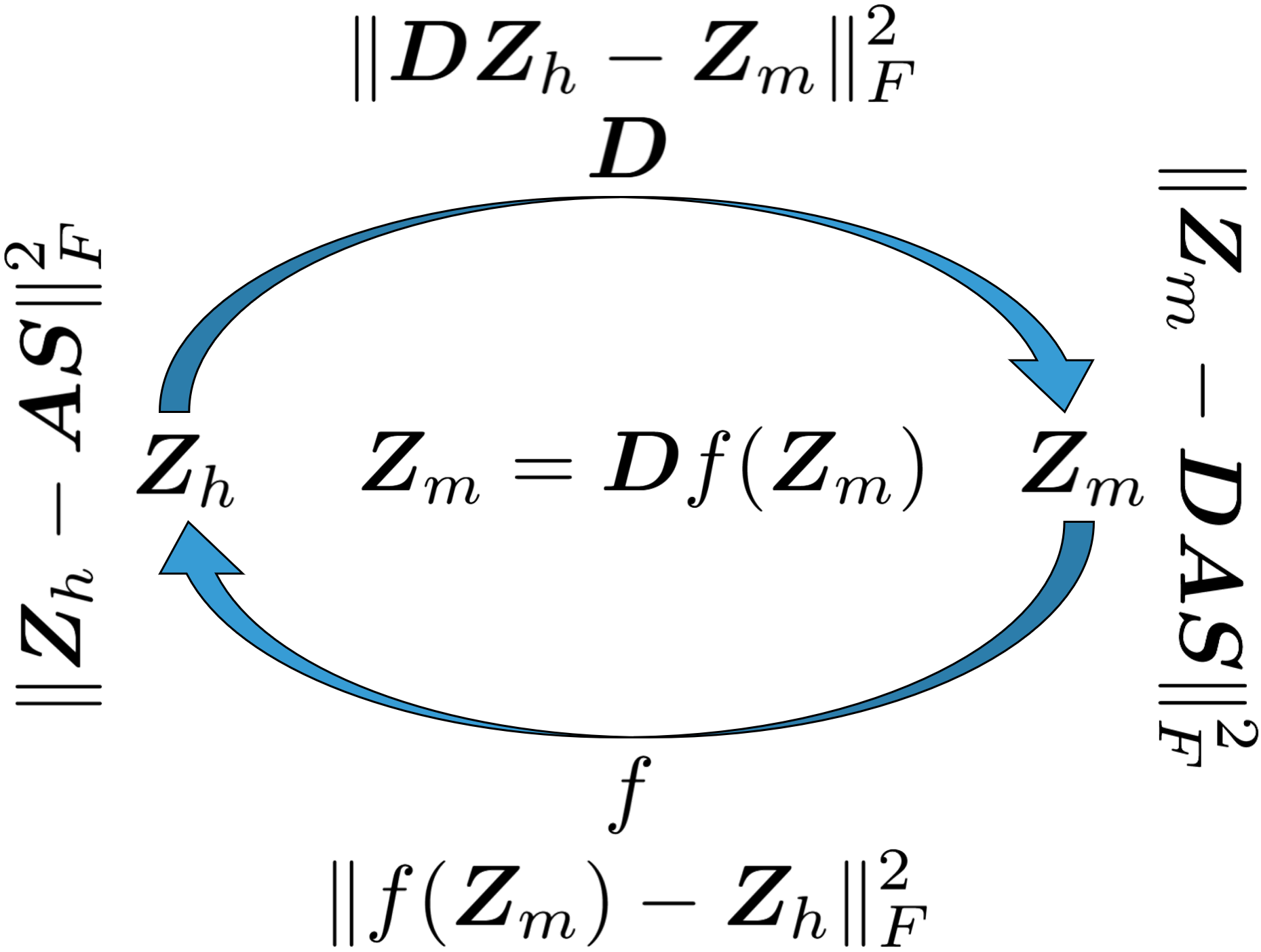}}
    \end{center}
    \vspace{-0.3cm}
    \caption{This graph illustrates the relations among the observable MSI $\bZ_m$, the virtual prism $f$ (to be implemented by the quantum deep network, QUEEN), the virtual HSI $\bZ_h$ generated by $f$, and the spectra response matrix $\bD$.
    This induces the data-fitting equation of $\bZ_m=\bD f(\bZ_m)$ with unknown $f$ to be optimized by the Adam optimizer.}
    \label{fig:ZhZmrelation}
    %\vspace{-0.4cm}
\end{figure}

A natural idea is to introduce an auxiliary variable that does not suffer from the underdetermined issue.
Specifically, we incorporate a virtual prism $f$ to further conduct light splitting on the observable optical information $\bZ_m$, thereby generating more virtual bands to form the auxiliary hyperspectral information $\bZ_h$.
This idea is technically feasible, though sounds impossible, as evidenced by very recent advances in the spectral super-resolution technique \cite{lin2022fast}.
To make the above idea clear, related quantities are graphically illustrated in Figure \ref{fig:ZhZmrelation}, from which a natural data-fitting criterion can be deduced as
$\|\bZ_m-\bD\bA\bS\|^2_F+
\|\bZ_h-\bA\bS\|^2_F+
\|f(\bZ_m)-\bZ_h\|^2_F+
\|\bD\bZ_h-\bZ_m\|^2_F$.
As we have $f(\bZ_m)=\bZ_h=\bA\bS$, the data-fitting term can be explicitly written as
\begin{align}\label{eq:DF}
\textrm{DF}(\bA,\bS,\bZ_h,f)
& \triangleq  
\|\bZ_m-\bD f(\bZ_m)\|^2_F+
\|\bZ_h-\bA\bS\|^2_F\notag\\
& ~~+
\|f(\bZ_m)-\bZ_h\|^2_F+
\|\bD\bZ_h-\bZ_m\|^2_F,
\end{align}
where, considering the high complexity of the function that the virtual prism should achieve, we introduce the rather advanced quantum feature extraction technique \cite{HyperQUEEN} to learn unitary features for realizing the function $f$ in \eqref{eq:DF}, as will be defined and detailed later.

Though \eqref{eq:DF} forming the cycling in Figure \ref{fig:ZhZmrelation} well resolves the underdetermined issue, the criterion itself is still ill-posed, and thus requires additional regularization.
Below, we discuss the regularization of $(\bA,\bS,\bZ_h,f)$.
First, the regularization of the hyperspectral signature matrix $\bA=[\ba_1,\dots,\ba_N]$ is tricky.
Accordingly to the well known Craig's criterion in the remote sensing area \cite{Craig1994}, the endmember simplex $\textrm{conv}\{\ba_1,\dots,\ba_N\}$ will hold a high resemblance to the minimum-volume data-enclosing simplex.
In fact, under mild practical conditions on the data purity, Craig's criterion has been proven to be a perfect conjecture with solid theoretical ground about two decades after it was proposed \cite{lin2015identifiability,lin2013EMI}.
This motivates us to regularize $\bA$ using the volume of the endmemeber simplex, i.e.,
\begin{equation}\label{eq:REGA}
V(\bA)\triangleq\textrm{volume}(\textrm{conv}\{\ba_1,\dots,\ba_N\}),
\end{equation}
where $\ba_i$ is the hyperspectral signature of the material $i$ (i.e., the $i$th endmember).

Moreover, the abundance maps $\bS$ are known to be sparse in practice \cite{patel2019abundance}, because each of its rows presents the spatial distribution of a particular material/substance over an often large area in remote sensing.
Such an abundance sparsity motivates us to regularize $\bS$ using the $\ell_1$-norm function, i.e,
\begin{equation}\label{eq:REGS}
    \|\bS\|_1\triangleq \sum_{i=1}^N\sum_{j=1}^L |S_{ij}|,
\end{equation}
where $|S_{ij}|$ denotes the absolute value of the $(i,j)$th entry of $\bS$.
As the $\ell_1$-norm function is known to be the convex envelope of the $\ell_0$-norm function (defined as the number of non-zero entries) \cite{CVXbookCLL2016}, it can promote the desired sparsity.

Note that as $\bA$ and $\bS$ are regularized, $\bZ_h=\bA\bS$ has been automatically regularized as well.
Thus, we only need to discuss the regularization of $f$, which is far more tricky.
Let $\mathcal{F}$ be the searching domain for the virtual prism $f$; in other words, $\mathcal{F}$ represents the space of all the possible functions $f$ that can be implemented by the deep network (i.e., Figure \ref{fig: networkv1}).
Since the network $f$ has to return a valid HSI $\bZ_h=f(\bZ_m)$, we should promote the spectral continuity and spatial continuity of the HSI, leading to the regularizer of $f$, i.e.,
\begin{equation}\label{eq:regf}
\textrm{REG}_f
\triangleq 
~\!\text{TV}_\text{spa}(f(\bZ_m))
+
\alpha~\!\text{TV}_\text{spe}(f(\bZ_m)),
\end{equation}
where $\alpha$ is the parameter controlling the relative regularization strength (empirically set as $\alpha:=1\textrm{E}-4$), and $\text{TV}_\text{spe}$ and $\text{TV}_\text{spa}$ are the spectral and spatial total-variation (TV) functions, respectively.
To define the two TV functions, $\text{TV}_\text{spe}$ is pixel-wise applied and defined as the $\ell_1$-norm of the gradient taken along the spectral dimension, and $\text{TV}_\text{spa}$ is band-wise applied and defined as the $\ell_1$-norm of the gradient taken along the spatial dimension.
These definitions can promote the desired continuity, because minimizing the $\ell_1$-norm promotes sparsity of the gradients, thereby encouraging the solution smoothness along the working dimension \cite{SUnSAL-TV}.
Instead of independently performing spectral super-resolution for each pixel, the $\text{TV}_\text{spa}$ function introduces a collaboration scheme among the pixels during the MU procedure.

As a summary, we propose to use the regularization function, i.e.,
\begin{equation}\label{eq:REGdef}
\textrm{REG}(\bA,\bS,f)
\triangleq 
V(\bA)+\|\bS\|_1+\textrm{REG}_f,
\end{equation}
leading to the overall MU criterion, i.e.,
\begin{align}\label{eq:MUcriterion}
\min_{\bA,\bS,\bZ_h,f\in\mathcal{F}}~&~\textrm{DF}(\bA,\bS,\bZ_h,f)+\lambda~\!\textrm{REG}(\bA,\bS,f)\notag\\
\text{s.t.}~~~~~~&~\bA\geq\mathbf{0},~\bS\geq\mathbf{0},~\bZ_h\geq\mathbf{0},
\end{align}
where $\lambda>0$ is the regularization parameter empirically set as $\lambda:=0.1$, and the three non-negativity constraints are oriented from the natures of the respective quantities.
As \eqref{eq:MUcriterion} is solved, we obtain the optimal $(\bA^\star,\bS^\star)$, from which we have the final MU solution; specifically, we have the multispectral endmember matrix $\bB^{\star}:=\bD\bA^{\star}$ and the abundance matrix $\bS^{\star}$ (cf. Algorithm \ref{alg: MUmain}).

\subsection{Geometry-Based Algorithm Design}\label{sec:GBAD}

Next, we design an algorithm to implement the proposed MU criterion \eqref{eq:MUcriterion}.
Directly handling it poses significant challenges, even if we are simply considering the ordering of the four block variables under the alternating optimization scheme.
%Directly handling it is too difficult even if we are simply talking about the optimization ordering of the four block variables.
%
Thus, we propose to merge $(\bA,\bS)$ into a single block variable, thereby allowing us to elegantly adopt convex geometry to simultaneously solve $(\bA,\bS)$, as will be detailed later.
Now, we just have three block variables, easier for us to decide the optimization ordering.
Considering the ease of the initialization, we propose to first optimize $f$, followed by optimizing $\bZ_h$, and finally the block variable $(\bA,\bS)$ is optimized, as specified in Algorithm \ref{alg: MUmain}.
We term our algorithm as prism-inspired multispectral endmember extraction (PRIME).

%====================================================
	\begin{algorithm}[t]
		\caption{PRIME Algorithm for Solving \eqref{eq:MUcriterion}}
		%----------------------------------------------------
		\begin{algorithmic}[1]\label{alg: MUmain}
            \STATE Given the MSI $\bZ_m$. Set $t:=0$.
            \STATE Initialize $\bZ_h^t$ by \eqref{eq: add noise}, and initialize $(\bA^t,\bS^t)$ by performing HU on $\bZ_h^t$ using HISUN \cite{HISUN}.
            
            \REPEAT
            \STATE Update $f^{t+1}$ by learning the relation between $\bZ_h^t$ and $\bZ_m$ using Adam optimizer \cite{Adam}.
		  
            \STATE Update $\bZ_h^{t+1}$ by \eqref{Zh closed form} using $\bA^t,\bS^t,f^{t+1}(\bZ_m)$, and $\bZ_m$.
            
            \STATE Update $(\bA^{t+1},\bS^{t+1})$ by processing $\bZ_h^{t+1}$ using HyperCSI \cite{HyperCSI}.
            
		  \STATE $t:=t+1$.
            \UNTIL the predefined stopping criterion is met.
            \STATE {\bf Output} multispectral endmember matrix $\bB^{\star}:=\bD\bA^t$ and abundance matrix $\bS^{\star}:=\bS^t$.
		\end{algorithmic}
		%----------------------------------------------------
	\end{algorithm}
%====================================================

With the light-splitting-driven algorithm initialization detailed in Appendix \ref{apx: init}, we now focus on deriving the solutions for the three block variables in the MU criterion \eqref{eq:MUcriterion}.
%
%The remaining task is to derive solutions for the three block variables in the MU criterion \eqref{eq:MUcriterion}.
%
According to \eqref{eq:DF}, \eqref{eq:regf}, \eqref{eq:REGdef}, and \eqref{eq:MUcriterion}, they amount to the following three subproblems, i.e.,
%
%%%%%% f %%%%%%
\begin{align}\label{update: f}
f^{t+1}\in\arg&\min_{f\in\mathcal{F}}~~
\|\bZ_m-\bD f(\bZ_m)\|^2_F+
\|f(\bZ_m)-\bZ_h^{t}\|^2_F\notag
\\
&+
\lambda(\text{TV}_\text{spa}(f(\bZ_m))
+
\alpha~\!\text{TV}_\text{spe}(f(\bZ_m))),
\end{align}
%%%%%%%%%%%%%%%
%
%%%%%% Z_h %%%%%%
\begin{align}\label{update: Z_h}
\bZ_h^{t+1}\in\arg\min_{\bZ_h\geq\mathbf{0}}~~
&
\|\bZ_h-\bA^{t}\bS^{t}\|^2_F+
\|f^{t+1}(\bZ_m)-\bZ_h\|^2_F\notag
\\
&+
\|\bD\bZ_h-\bZ_m\|^2_F,
\end{align}
%%%%%%%%%%%%%%%
%
%%%%%% AS %%%%%%
\begin{align}\label{update: AS}
\left(\bA,\bS\right)^{t+1}\in\arg\min_{\bA,\bS\geq\mathbf{0}}
&
~\frac{\|\bZ_h^{t+1}-\bA\bS\|^2_F}{\lambda}+
V(\bA)+\|\bS\|_1,
\end{align}
%%%%%%%%%%%%%%%
%
which will be alternatively updated in Algorithm \ref{alg: MUmain}.
To solve \eqref{update: f}, we design $f$ as a QUEEN-based quantum prism network, and train the network to learn the light-splitting relation between $\bZ_h^t$ and $\bZ_m$ using adaptive moment estimation (Adam) optimizer \cite{Adam}.
The detailed rationale behind the design of $f$ will be thoroughly discussed in Section \ref{sec:fDesign}, where the objective function in \eqref{update: f} will elegantly serve as the loss function during the model training procedure.
%As there is a long story behind the network design of $f$, it will be collectively presented in Section \ref{sec:fDesign}, where \eqref{update: f} will be considered as the loss function during the training phase.

To solve \eqref{update: Z_h}, the non-negative constraint would make the optimization slow.
Considering that the overall criterion \eqref{eq:MUcriterion} is non-convex, we may not need to pursue a global optimal solution.
Thus, for efficient optimization, we solve the objective function in \eqref{update: Z_h}, followed by projecting the solution to the non-negative orthant to catch up with the constraint of \eqref{update: Z_h}, leading to the update of 
\begin{align}
\bZ_h^{t+1} = \Pi ( 
( 2 \bI_{M} + \bD^T \bD )^{-1} 
( \bA^t \bS^t +  f^{t+1}(\bZ_m) +  \bD^T \bZ_m ) 
),\label{Zh closed form}
\end{align}
where $\Pi(\cdot)$ is the projector onto the non-negative orthant.
%, as presented in Algorithm \ref{alg: MUmain}.

To solve \eqref{update: AS}, note that even without the regularizers $V(\bA)$ and $\|\bS\|_1$, the first term of \eqref{update: AS} is already the NP-hard NMF problem.
That said, we need to find a strategy to efficiently address \eqref{update: AS}.
Our idea is to employ the convex geometry algorithm called Hyperplane-based Craig Simplex Identification (HyperCSI) \cite{HyperCSI}, which is not only extremely fast but also exactly matches the physical meaning of \eqref{update: AS}, as illustrated below.
Let us explain how the three terms in \eqref{update: AS} associate with the mechanisms in HyperCSI.
For the first term, the problem ``$\min_{\bA,\bS\geq\mathbf{0}}~\!\frac{\|\bZ_h^{t+1}-\bA\bS\|^2_F}{\lambda}
~\equiv~
\min_{\bA,\bS\geq\mathbf{0}}~\!\|\bZ_h^{t+1}-\bA\bS\|^2_F$" (cf. $\lambda>0$) is exactly the unregularized NMF criterion for performing HU on $\bZ_h^{t+1}$ \cite{NMF-HU,HyperCSI}, and HyperCSI is exactly proposed for high-performance HU.
For the second term $V(\bA)$, it encourages a solution $\bA=[\ba_1,\dots,\ba_N]$ to form the minimum-volume simplex $\textrm{conv}\{\ba_1,\dots,\ba_N\}$ (cf. \eqref{eq:REGA}), and this is exactly what Craig's criterion proposes to do \cite{Craig1994}.
This minimum-volume mechanism is exactly covered by HyperCSI, because HyperCSI is developed based on Craig's minimum-volume HU criterion \cite[Equation (8)]{HyperCSI}.
For the third term $\|\bS\|_1$, this is to promote the abundance sparsity of $\bS$, as discussed in \eqref{eq:REGS}.
Notice that in HyperCSI, instead of estimating the vertices of the minimum simplex, it equivalently estimates the hyperplanes forming the boundary of the minimum simplex.
Using such a hyperplane geometry is exactly to capture the abundance sparsity of $\bS$, as detailed in \cite[Section III-E]{HyperCSI}, where the key theory is that $[\bS]_{(i,n)}=0$ (sparsity) if, and only if, the $n$th pixel of $\bZ_h^{t+1}$ belongs to the $i$th hyperplane $\mathcal{H}_i\triangleq\textrm{aff}(\{\ba_1,\dots,\ba_N\}\setminus\{\ba_i\})$ \cite[Section III-E]{HyperCSI}.
Thus, identifying the boundary hyperplanes also promotes the abundance sparsity.
Overall, both \eqref{update: AS} and HyperCSI are to conduct HU using the minimum-volume and sparsity regularization mechanisms.
Therefore, we update $(\bA^{t+1},\bS^{t+1})$ of \eqref{update: AS} by processing $\bZ_h^{t+1}$ using HyperCSI \cite{HyperCSI}, whose compression and radius parameters are set as $\eta:=1$ and $r:=1\textrm{E}-8$, respectively.
We have completed the design of the PRIME algorithm for MU, which is summarized in Algorithm \ref{alg: MUmain}.

To conclude this section, we remark that PRIME judiciously blends the advantages of both convex analysis (CO) and deep learning (DE), echoing the recent trend in remote sensing and imaging areas \cite{ADMM-ADAM}.
This CODE theory has spurred numerous latest remote sensing technologies, such as mangrove mapping \cite{CODEMM}, image fusion \cite{CODEIF}, and change detection\cite{CODEHCD}, etc.

\subsection{Quantum Prism Design}\label{sec:fDesign}

Let us specify how we update $f$ in \eqref{update: f}, whose architecture design and training strategy will be discussed in this section.
If we could have a pretrained $f$, then the alternating step \eqref{update: f} could actually be omitted.
However, the mission of $f$ is to perform the virtual light-splitting function, which does not associate with a real HSI sensor; that said, big data collection for pretraining $f$ does not seem to be feasible, due to the lack of ground-truth real HSI.
This motivates us to train $f$ using the deep image prior (DIP) strategy \cite{DIP2018}, because this can be done by using just the single data (i.e., the observable MSI $\bZ_m$).
Another great advantage is that DIP is an unsupervised strategy, which can be done in a blind manner.

\begin{table}[t]
\scriptsize
\setlength{\tabcolsep}{4pt} % Default value: 6pt
\caption{Quantum gates used in the quantum prism $f$, their symbols, and the corresponding unitary operators. In the mathematical definitions, let $\delta \triangleq \cos(\theta/2)$ and $\zeta \triangleq \sin(\theta/2)$.}\label{tab: common_qu_gate}
\vspace{-0.3cm}
\begin{center}
\begin{tabular}{|c c c|} 
 \hline
 \rule{0pt}{2ex}
 Quantum Gate & Symbol & Unitary Operator 
 \rule{0pt}{2ex}
 \\
 \hline
 \rule{0pt}{4ex}
 Rotation X
 &
 \begin{tikzcd}
    \qw & \gate{R_{X}(\theta)} & \qw
 \end{tikzcd}
 &
 $\begin{pmatrix}
    \delta & -i \zeta \\
    -i \zeta & \delta
\end{pmatrix}$
 \rule{0pt}{4ex}
 \\ 
 \hline
 \rule{0pt}{4ex}
 Rotation Y
 &
 \begin{tikzcd}
    \qw & \gate{R_{Y}(\theta)} & \qw 
 \end{tikzcd}
 & 
 $\begin{pmatrix}
    \delta & - \zeta \\
    \zeta & \delta
\end{pmatrix}$
 \rule{0pt}{4ex}
\\
 \hline
 \rule{0pt}{4ex}
 \rule{0pt}{6.5ex}
 Ising XX
&
\begin{quantikz}
    \qw & \gate{XX(\theta)} & \qw
\end{quantikz}
&
$\begin{pmatrix}
    \delta & 0 & 0 & -i\zeta\\
    0 & \delta & -i\zeta & 0\\
    0 & -i\zeta & \delta & 0\\
    -i\zeta & 0 & 0 & \delta
\end{pmatrix}$
\\
\hline
Pauli-Z &
\begin{tikzcd}
\meter{} 
\end{tikzcd}
&
$\begin{pmatrix}
    1 & 0 \\
    0 & -1 \\
\end{pmatrix}$
%\rule[-5ex]{0pt}{4ex}
 \\
 \hline
NOT &
\begin{tikzcd}
\qw & \gate{X} & \qw
\end{tikzcd}
&
$\begin{pmatrix}
    0 & 1 \\
    1 & 0 \\
\end{pmatrix}$
%\rule[-5ex]{0pt}{4ex}
 \\
 \hline
 \rule{0pt}{11ex}
Toffoli (CCNOT)
 &
 \begin{tikzcd}
    \qw & \ctrl{1} & \qw \\
    \qw & \octrl{1} & \qw \\
    \qw & \targ{} & \qw
 \end{tikzcd}
 &
 $\textrm{DIAG}(\bm{I}_4,X,\bm{I}_2)$
 \rule[-5ex]{0pt}{4ex}
 \\
 \hline
\end{tabular}
 \vspace{-0.7cm}
\end{center}
\end{table}

To be more specific, the network $f$ accepts the input $\bZ_m$, and outputs the virtual HSI $\bZ_h$, merely from which we can unsupervisedly train/optimize the network parameters with the loss function provided by \eqref{update: f}.
Note that through the training/optimization of \eqref{update: f}, the single data pair $(\bZ_m,\bZ_h^t)$ is sufficient to complete the learning of the light-splitting relation.
Since both the training data (i.e., $\bZ_m$) and network architecture $f$ (to be designed next) are fixed, we do not need to train $f^{t+1}$ from scratch.
Instead, we can initialize the network parameters as $f^t$ when training $f^{t+1}$ during the $(t+1)$th iteration of Algorithm \ref{alg: MUmain}.
In other words, there are $10$ iterations in PRIME, where $f$ is carefully trained with $100$ epochs during the first iteration, while $f$ is fine-tuned with only $30$ epochs in the remaining nine iterations.
%
%Though the training data and network architecture are both fixed, the training in the $(t+1)$th iteration still contributes to better MU results.
%
As the PRIME algorithm evolves as the iteration number $t$ increases, the virtual HSI $\bZ_h^t$ gets better, and through the criterion \eqref{update: f}, this better information of $\bZ_h^t$ is injected into the learning of $f$, thereby obtaining a better light-splitting function $f^{t+1}$.
In this work, \eqref{update: f} is optimized via the Adam optimizer (learning rate set as $0.005$) \cite{Adam}.

    \begin{figure*}[ht]
    \begin{center}
    \resizebox{0.95\linewidth}{!}{\hspace{-0cm}\includegraphics{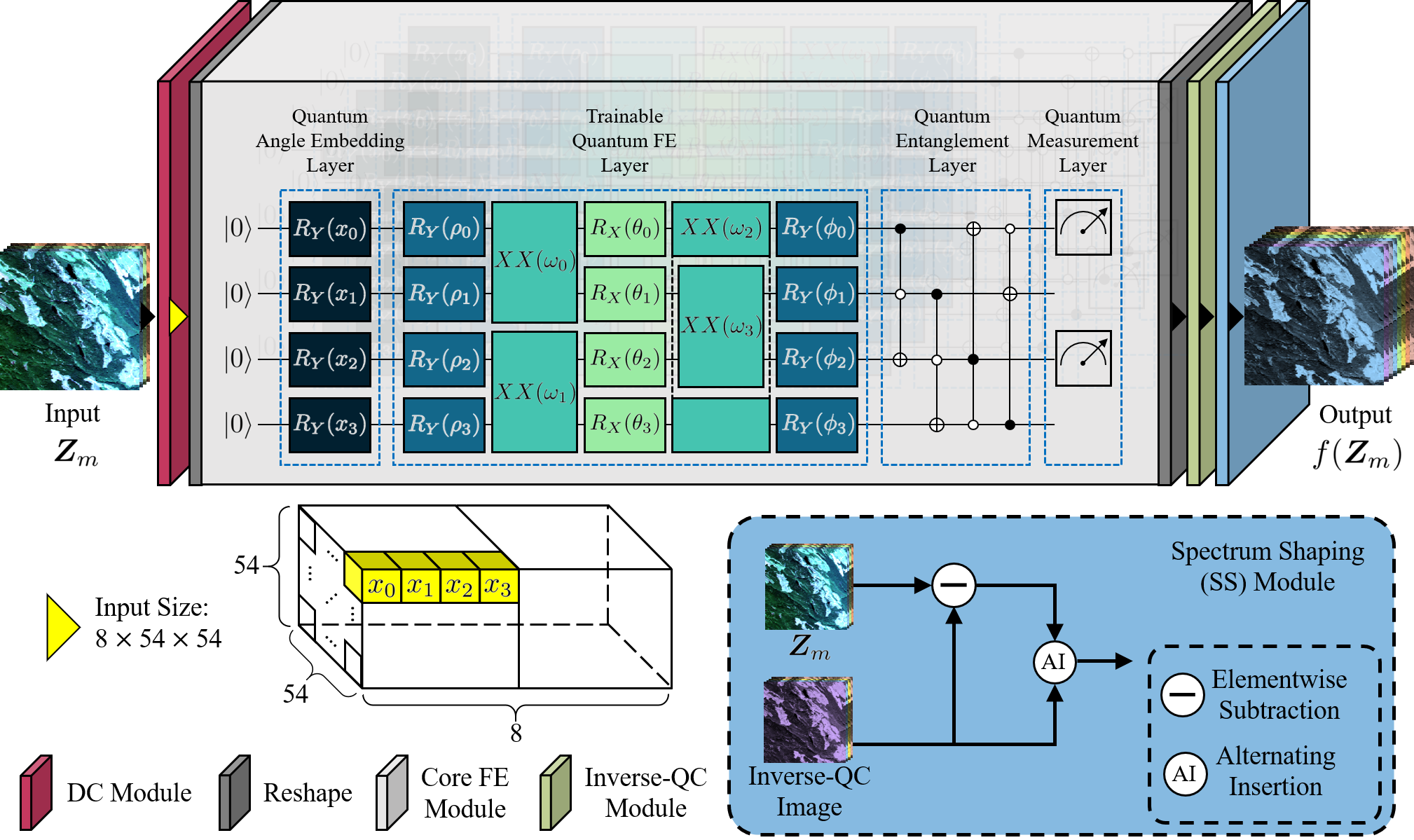}}
    \end{center}
    \vspace{-0.3cm}
    \caption{The illustrating architecture of the QUEEN-based \cite{HyperQUEEN} quantum prism network $f$, where the learning parameters include $(\rho_i,\omega_i,\theta_i,\phi_i)$.
    Note that $x_i$'s are not learning parameters but for data encoding, meaning that the feature value $x_i$ is encoded as the angle of the quantum rotation gate $R_Y$ (i.e., quantum angle embedding).
    DC module highly compresses the input image information into very few qubits, which are then fed into the core FE module with the cross-feature ordering specified in the yellow cubes.
    Provably, the core FE module can realize any valid quantum operators.
    Once the core quantum signal processings are done with the core FE module, we use the inverse-QC module to read out the well processed quantum signals (cf. \cite[Section II-C]{HyperQUEEN}).
    Finally, the SS module is designed according to the light-splitting equation of $0.5([\bZ_m]_{(i,:)}\pm [{\bm \theta}]_{(i,:)})$ (detailed in Appendix \ref{apx: init}) for shaping the pixel spectra.
    }
    \label{fig: networkv1}
    %\vspace{-0.4cm}
    \end{figure*}

Next, we design the QUEEN-based prism function $f$.
We will adopt the quantum feature extraction technique developed in HyperQUEEN \cite{HyperQUEEN}, partly because we believe that the advanced concept of quantum features can contribute to the learning of the complicated light-splitting task (i.e., spectral super-resolution), and partly because the low-rank module in HyperQUEEN may be interpreted as the spectral super-resolution module (cf. \cite[Table II]{HyperQUEEN}).
To get a sense, QUEEN is a deep network composed of quantum neurons, as its name suggested.
Here, quantum neurons are commonly referred to as quantum logic gates.
The critical ones used in $f$ are summarized in Table \ref{tab: common_qu_gate}, including the rotation gates, Ising gate, Pauli gate, and Toffoli gate, etc.
%Some frequently used ones are summarized in Table \ref{tab: common_qu_gate}, including the rotation gates, Ising gate, Pauli gate, and Toffoli gate, etc.
%
One can see that all these quantum gates correspond to some unitary operators \cite[Section II]{HyperQUEEN}, defined in Table \ref{tab: common_qu_gate}, where we also display the symbols of these gates.

%The design of the quantum prism network $f$ is graphically illustrated in Figure \ref{fig: networkv1}, which is composed of four modules, including the deep compression (DC) module, the core full expressibility (FE) module, the spectrum shaping (SS) module, and the inverse-QC module for addressing the quantum collapse (QC) effect, to be introduced below.
The design of the quantum prism network $f$ is graphically illustrated in Figure \ref{fig: networkv1}, which is composed of four modules, including the deep compression (DC) module, the core full expressibility (FE) module, the spectrum shaping (SS) module, and the inverse-QC module for addressing the quantum collapse (QC) effect.
To explicitly describe the network architecture for reproducibility while maintaining the readability of the paper, we provide the network details (e.g., configuration and output size of each layer) in Appendix \ref{apx: network}.
The DC module is proposed to account for the very limited quantum bit (qubit) resources in current quantum computers; for example, the almost most advanced quantum computer called ``IBM Osprey" has only 433 qubits.
Thus, the DC module (defined in Appendix \ref{apx: network}) highly compresses the image information into very few qubits, for conducting the core quantum image processing in the highly compressed feature space \cite{HyperQUEEN}.
The features $x_i$'s are then fed into the core FE module as the angles of the quantum rotation gates $R_Y$ (i.e., quantum angle embedding \cite{QAEBD}) by using the cross-feature ordering specified in the yellow cubes of Figure \ref{fig: networkv1}.
This cross-feature strategy is to allow more different features to interact with each other, thereby having a deeper understanding of related information \cite{HyperQUEEN}.
The core FE module is more complicated, so we introduce it later.

Once the core quantum signal processings are done with the core FE module, we use the inverse-QC module (defined in Appendix \ref{apx: network}) to read out the well processed quantum signals; the related theory can be found in \cite[Section II-C]{HyperQUEEN}.
Simply speaking, the QC effect says that once we attempt to read/observe the quantum state of the processed signals, it will immediately collapse to some eigenstate; thus, the function of the inverse-QC module aims at learning the inverse mapping from some measuring statistics (associated with the collapsed quantum state) to the target image \cite{HyperQUEEN}.
On the other hand, the SS module (defined in Appendix \ref{apx: network}) is designed according to the aforementioned light-splitting relation that splits an MSI band into $\gamma:=2$ HSI bands ($\gamma$ is the upsampling factor in the spectral super-resolution task), i.e., $0.5([\bZ_m]_{(i,:)}\pm [{\bm \theta}]_{(i,:)})$ (cf. Appendix \ref{apx: init}).
This relation allows us to much more efficiently focus on learning only half of the bands (e.g., the odd bands), followed by generating the remaining even bands for alternately inserting in between the odd bands, thereby generating the complete virtual HSI $\bZ_h=f(\bZ_m)$, as graphically illustrated in the lower-right block of Figure \ref{fig: networkv1}.

To well process the quantum signals, we need a strong core processing module, such as a module with FE \cite[Theorem 2]{HyperQUEEN}.
Therefore, we employ a recently developed Ising-Rotation quantum network architecture to design our core quantum FE module, sequentially composed of Rotation Y gate, Ising XX gate, Rotation X gate, Ising XX gate, and Rotation Y gate.
We then have the mathematical guarantee that the core FE module can express any valid quantum operators, as stated in the following theorem.
\begin{theorem} \label{theorem: FE}
The trainable quantum neurons deployed in the core FE module of the proposed quantum prism network $f$ (cf. Figure \ref{fig: networkv1}) can express any valid quantum unitary operator $U$, with some real-valued trainable network parameters $\{\rho_{k},\omega_{k},\theta_{k},\phi_{k}\}$.
\hfill$\square$
\end{theorem}
The proof exactly follows the procedure of proving \cite[Theorem 2]{HyperQUEEN}, and is omitted here for conciseness.
As also proved in \cite{HyperQUEEN}, the Toffoli quantum entanglement is useful, and hence utilized in our core FE module as well.
In the last stage, the Pauli-Z gate is used to measure the quantum information using a strategy similar to the max pooling, and such a strategy is also proven useful even in some rather complicated inverse imaging tasks \cite{HyperQUEEN}.
Overall, the core FE module is defined in Appendix \ref{apx: network}.
Therefore, we have completed the design of the QUEEN-based quantum prism network $f$.

The interested readers are referred to \cite[Section II-A]{HyperQUEEN} and \cite[Section II-B]{HyperQUEEN}, where a comprehensive review of the fundamental concepts required in QUEEN is presented, including quantum deep learning, barren plateaus effect, Dirac notation system, and quantum measurement/collapse, etc.
Next, we experimentally demonstrate the strength of this QUEEN-based PRIME algorithm for MU.

%In the DC module and the inverse-QC module, the $3\times3$ convolution and the $3\times3$ transposed convolution are added to network layers to exploit the spatial pixel collaboration instead of independently performing spectral super-resolution for each pixel.

\begin{figure}[ht]
    \begin{center}
    \resizebox{0.95\linewidth}{!}{\hspace{-0.1cm}\includegraphics{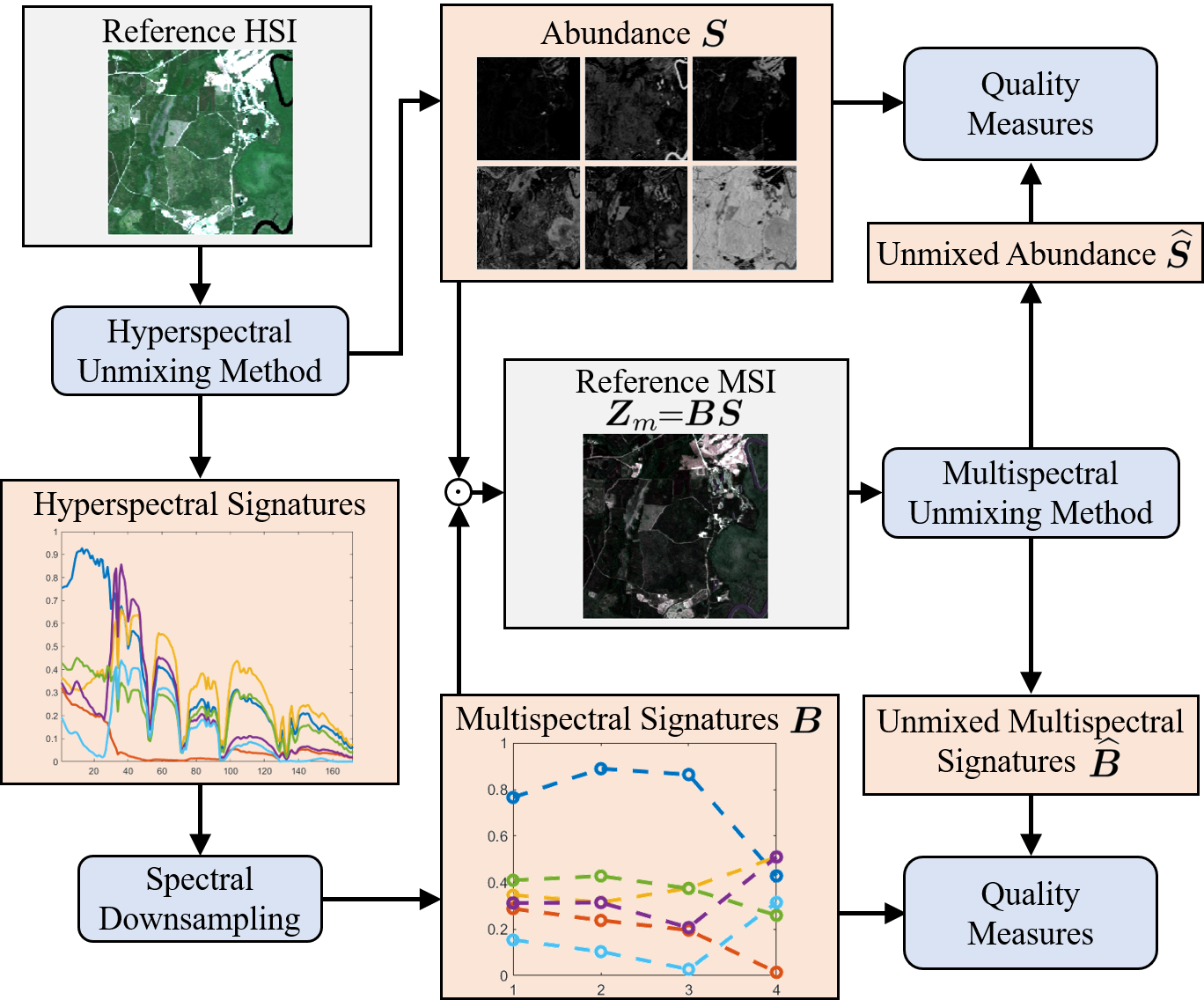}}
    \end{center}
    \vspace{-0.3cm}
    \caption{Flowchart of the proposed Lin’s protocol for evaluating the performance of MU methods.
    }
    \label{fig: protocol}
    %\vspace{-0.4cm}
\end{figure}

\begin{figure}[t]
    \begin{center}
    \resizebox{0.98\linewidth}{!}{\hspace{-0.1cm}\includegraphics{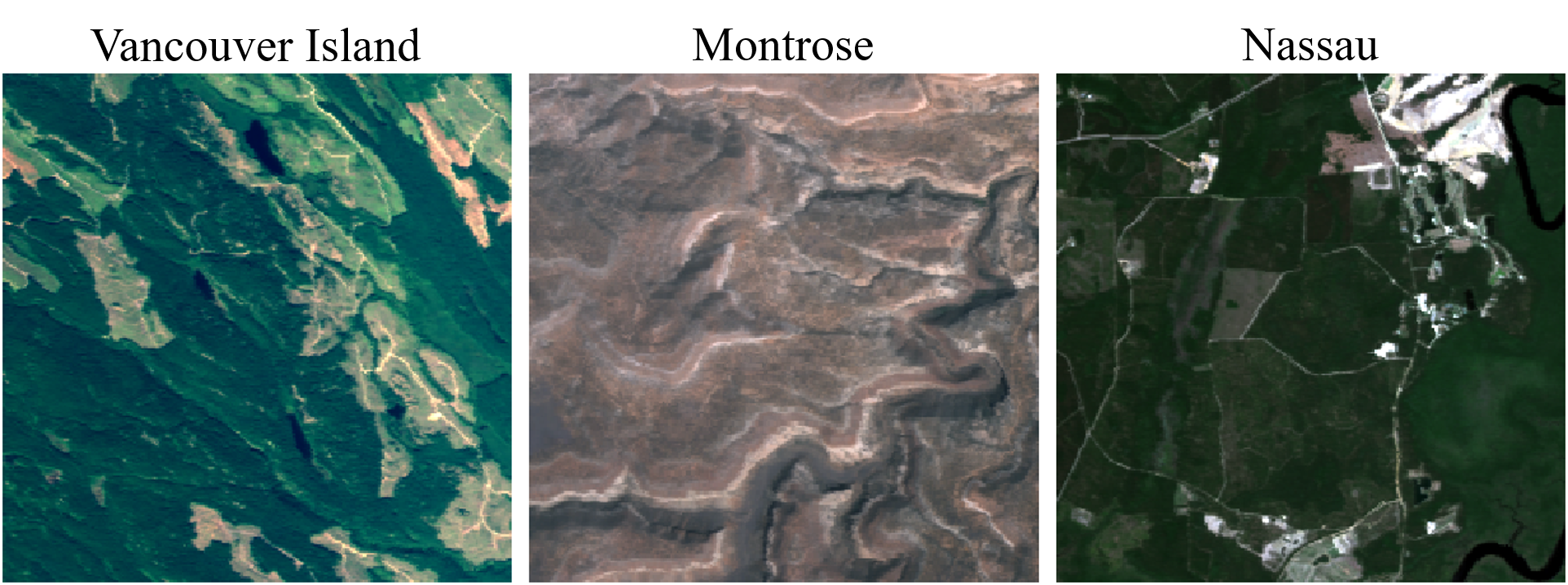}}
    \end{center}
    \vspace{-0.3cm}
    \caption{False-color compositions of the investigated data acquired over different studied regions.}
    \label{fig: GT}
    %\vspace{-0.4cm}
\end{figure}

\section{Experimental Results}\label{sec: experiment}

This section aims to demonstrate the effectiveness of the proposed PRIME algorithm.
As there is no MU work targeted at the underdetermined case, we propose an experimental protocol for evaluating the performance of the MU algorithms in Section \ref{sec: exp_design}, where one can also find the detailed experimental setting and data description.
In Section \ref{sec: qqcompare}, we conduct comprehensive comparisons with the MU methods used in \cite{MU1}.
The ablation study for the PRIME algorithm is investigated in Section \ref{sec: ablation}.
Finally, in Section \ref{sec: discussion}, we show that the proposed PRIME algorithm is generally applicable to different sensors, and analyze the impact of the upsampling factor $\gamma$ on the performance of PRIME.

\subsection{Experimental Protocol and Design}\label{sec: exp_design}

Since prior MU works do not consider the underdetermined scenario, we cannot find suitable experimental protocols to evaluate the MU performances.
In view of this, we design a protocol (cf. Figure \ref{fig: protocol}) to facilitate future development of the critical MU technique.
Since there is no benchmark MU dataset with available ground-truth (GT) $\bB$ and $\bS$, our protocol (referred to as Lin's protocol) is particularly designed to address this issue.
This protocol is inspired by Wald's protocol, which has been widely used for MSI/HSI pansharpening quality evaluation \cite{Wald,9245579,7284770,10804644}.
Specifically, we adopt the reduced resolution (RR) assessment \cite[Figure 2]{7284770}, which mirrors the synthesis property of Wald's protocol.
Specifically, the process begins with a reference HSI, from which effective HU methods (e.g., HISUN) are used to first generate the GT abundances $\bS$ and reference hyperspectral signatures $\bA_\textrm{ref}$, where $\bA_\textrm{ref}$ is further uniformly downsampled along the spectral dimension to obtain the GT multispectral signatures (endmembers) in $\bB$.
Next, $\bS$ and $\bB$ are used to synthesize the reference MSI $\bZ_m=\bB\bS$, which are then fed into the MU methods to estimate the signatures $\widehat{\bB}$ and abundances $\widehat{\bS}$.
Finally, the similarity between $\widehat{\bS}$ and ${\bS}$, as well as the similarity between $\widehat{\bB}$ and ${\bB}$, are measured with suitable quality metrics, as illustrated in Figure \ref{fig: protocol}.

\begin{figure*}[t]
    \begin{center}
    \resizebox{0.95\linewidth}{!}{\hspace{-0.1cm}\includegraphics{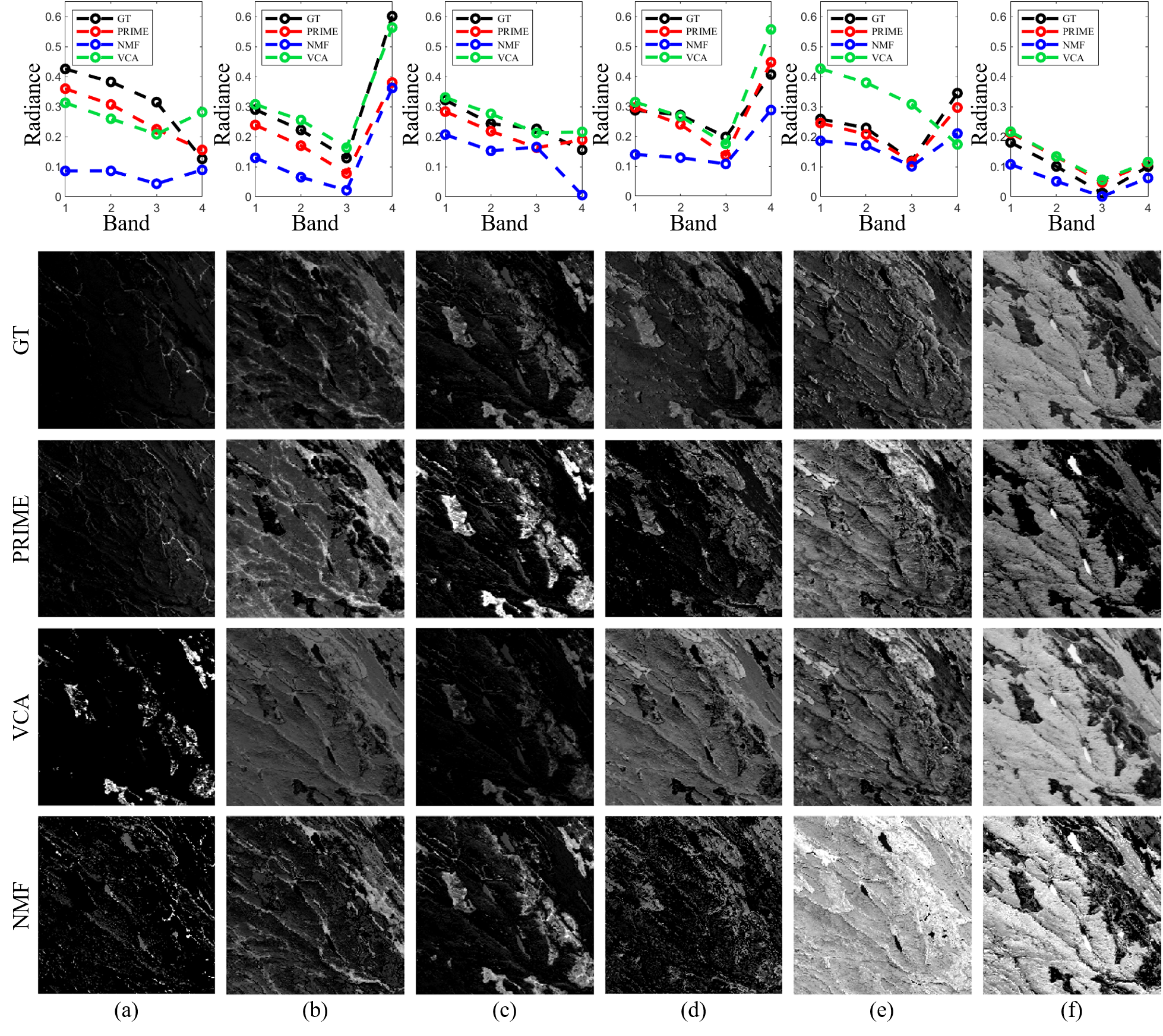}}
    \end{center}
    \vspace{-0.3cm}
    \caption{The MU results, including the estimated abundance maps and the multispectral endmember signatures, for the Vancouver Island data.}
    \label{fig: Vancouver Island}
    \vspace{-0.4cm}
\end{figure*}

For the reference HSI to be used in the protocol, we consider the data acquired by NASA’s Airborne Visible/Infrared Imaging Spectrometer (AVIRIS) sensor \cite{AVIRISdata}.
The AVIRIS sensor is composed of $224$ contiguous bands, and only $172$ bands are used in our experiments after removing those water-vapor absorption bands (i.e., bands 1-10, 104-116, 152-170, and 215-224).
Then, the spectral downsampling stage in the protocol is implemented using a uniform spectral downsampling matrix, which is widely accepted in the remote sensing area \cite{COCNMF,lin2023metasurface}.
This approach effectively and fairly merges spectral information across specified wavelength ranges to generate representative bands.
Precisely, the spectral ranges correspond to the Landsat Thematic Mapper (TM) bands 1–4, covering 450–520, 520–600, 630–690, and 760–900 nm, respectively, are used to uniformly downsample the reference signatures $\bA_\text{ref}$, resulting in the $4$-band GT multispectral signatures $\bB$.
%Then, the spectral ranges of the spectral downsampling stage in the protocol correspond to the Landsat TM bands 1–4 (covering 450–520, 520–600, 630–690, and 760–900 nm regions, respectively), in order to spectrally downsample the reference signatures $\bA_\textrm{ref}$ to obtain the $4$-band GT multispectral signatures $\bB$ (i.e., $P=4$).}
%
Three data from diverse regions are introduced below.
The first data (with a spatial resolution of 17.3 m) was acquired over Vancouver Island, British Columbia, Canada.
The second data (with a spatial resolution of 11.7 m) was acquired over Montrose, CO, USA.
The third data (with a spatial resolution of 16.8 m) was acquired over Nassau, FL, USA.
All the three reference images are with $L=256\times256$ pixels, and their false-color images (composed of band 20 (R), band 12 (G), and band 6 (B)) are displayed in Figure \ref{fig: GT}.

All the experiments are conducted under Mathworks MATLAB R2023b and PyTorch 3.9 on a personal computer equipped with an Intel Core-i9-10900X CPU with a speed of 3.70-GHz and 64 GB RAM, and two NVIDIA RTX-2080Ti GPUs.
To demonstrate the efficacy of PRIME over diverse data regions, two popular metrics \cite{7284770,7010915,COCNMF,CODEIF} are used in the experiment for quantitatively evaluating the MU performance.
In particular, we use the spectral angle mapper (SAM) for evaluating the similarity between the reference signatures $\bB$ and estimated signatures $\widehat{\bB}$, defined as
\begin{align}
    \text{SAM}=\frac{1}{N}\sum_{n=1}^{N}\text{arccos}\left(\frac{\left([\widehat{\bB}]_{(:,n)}\right)^T\left([\bB]_{(:,n)}\right)}{\| [\widehat{\bB}]_{(:,n)} \|_2 \cdot \| [\bB]_{(:,n)} \|_2}\right).
\end{align}
Also, the root-mean-squared error (RMSE), defined as $\frac{1}{\sqrt{NL}}\|\widehat{\bS}-\bS\|_F$, is employed to quantitatively measure the similarity between the reference abundance $\bS$ and estimated abundance $\widehat{\bS}$.
Smaller values of SAM/RMSE indicate better MU performances.

\begin{figure*}[t]
    \begin{center}
    \resizebox{0.95\linewidth}{!}{\hspace{-0.1cm}\includegraphics{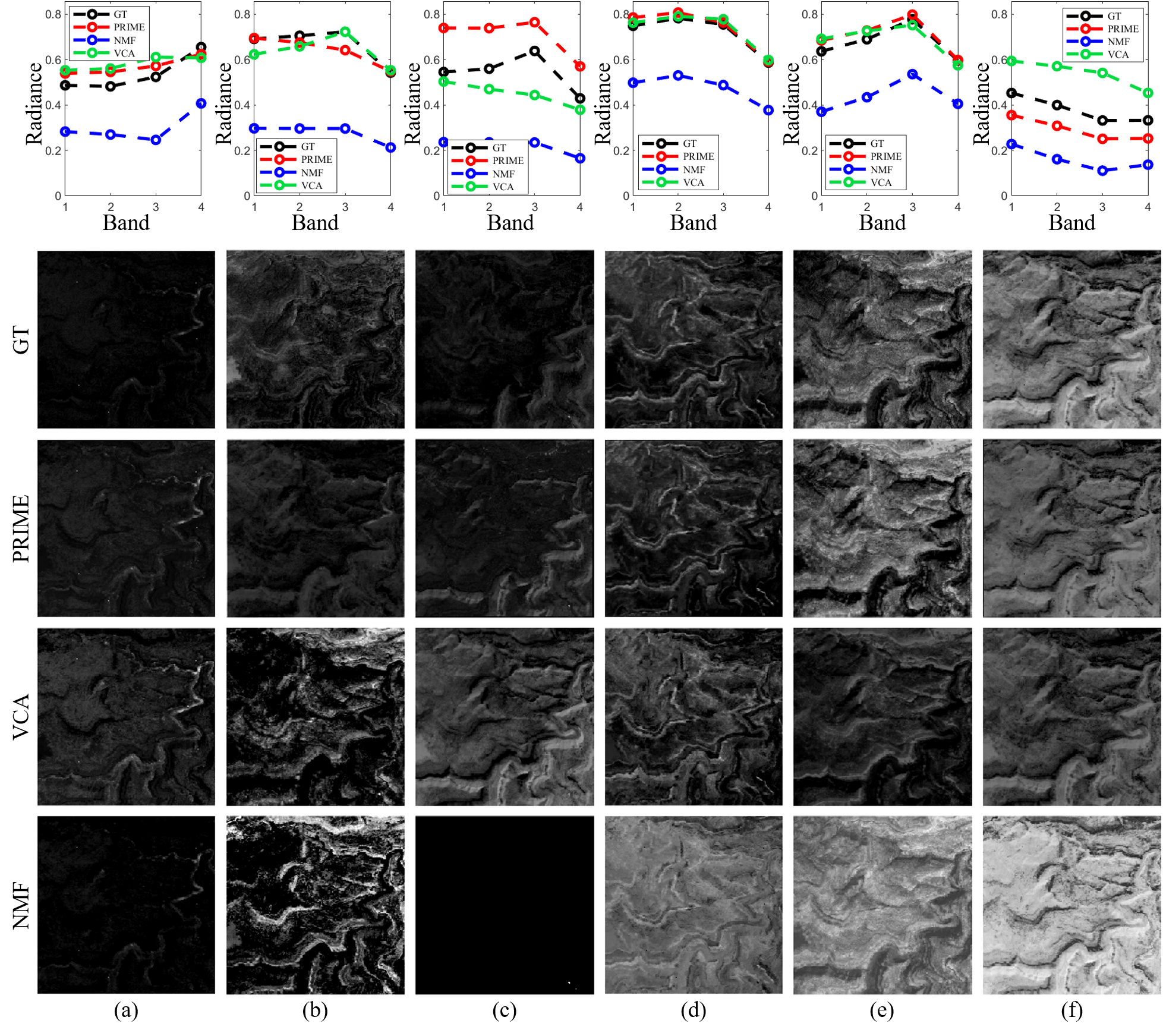}}
    \end{center}
    \vspace{-0.3cm}
    \caption{The MU results, including the estimated abundance maps and the multispectral endmember signatures, for the Montrose data.}
    \label{fig: Montrose}
    \vspace{-0.4cm}
\end{figure*}

\begin{figure*}[t]
    \begin{center}
    \resizebox{0.95\linewidth}{!}{\hspace{-0.1cm}\includegraphics{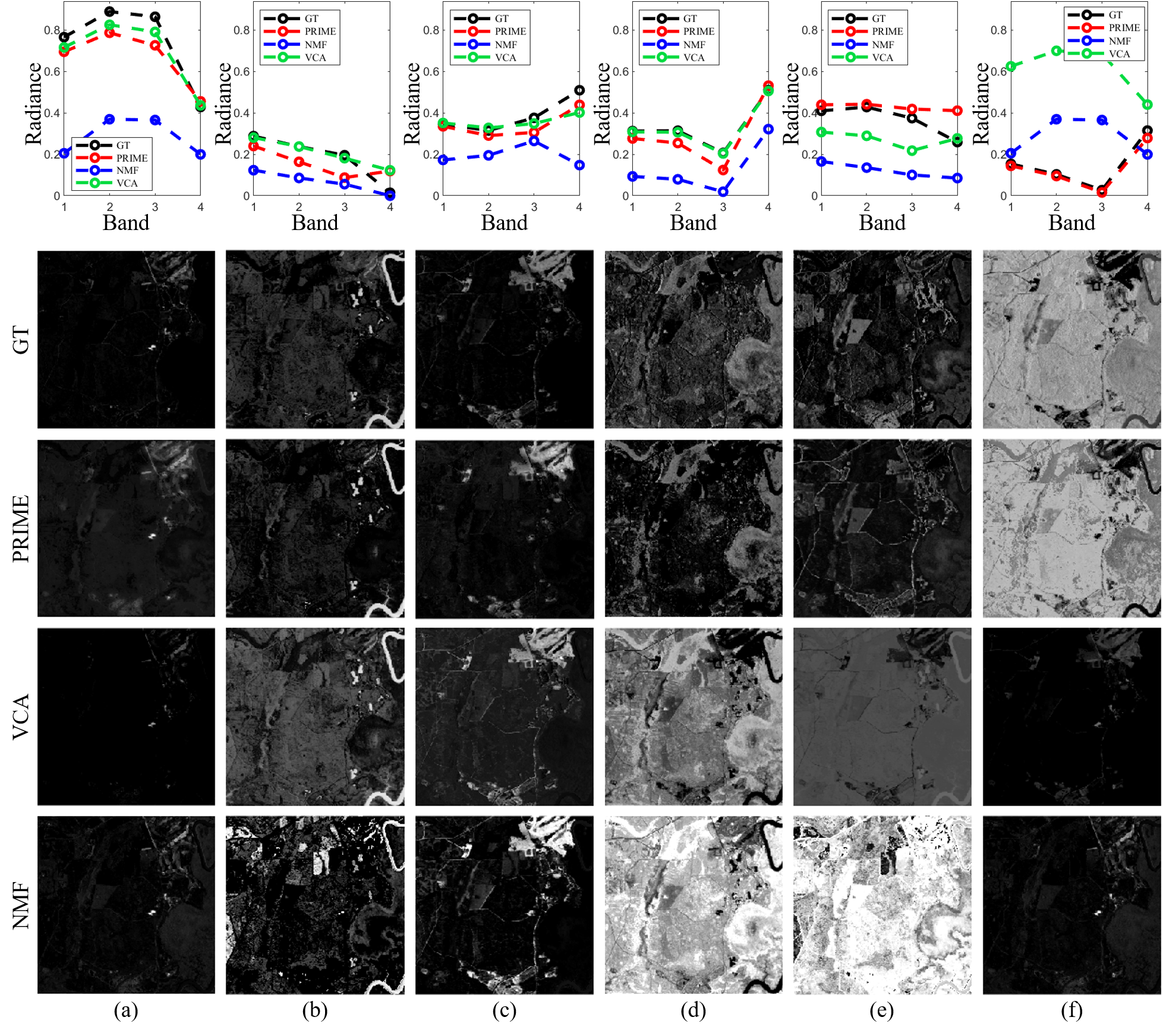}}
    \end{center}
    \vspace{-0.3cm}
    \caption{The MU results, including the estimated abundance maps and the multispectral endmember signatures, for the Nassau data.}
    \label{fig: Nassau}
    \vspace{-0.4cm}
\end{figure*}

\subsection{Qualitative and Quantitative Analysis}\label{sec: qqcompare}

In this section, we conduct qualitative and quantitative analysis for the proposed PRIME algorithm (i.e., Algorithm \ref{alg: MUmain}) together with related methods, on the three $4$-band MSI data described in Section \ref{sec: exp_design} (i.e., $P:=4$).
Also, we empirically set $N:=6$ for all the three data.
There are rather few existing MU methods.
We will compare with the MU methods used in \cite{MU1}, which extends VCA \cite{VCA} and NMF \cite{NMF-peer} to conduct the MU task.
Note that VCA will first perform dimensionality reduction (DR). 
Considering that the number of bands is smaller than the sources in MU (i.e., $P<N$), we ignore this DR step by running the algorithm in the $P$-dimensional space.
Since VCA can only compute the signatures $\bB$, we compute the abundance $\bS$ by left-multiplying $\bZ_m$ with the pseudo-inverse of $\bB$, as suggested by the original paper \cite{VCA}.
As for NMF, the iteration number is set to $1000$ for better convergence.
Besides, $\bB$ and $\bS$ in NMF are initialized by the powerful HISUN as a warm start strategy \cite{HISUN}.
As NMF itself cannot yield a good abundance estimation result, we further refine $\bS$ by solving the non-negative least-squares problem using the optimized $\bB$ returned from NMF, in order to upgrade the performance of NMF.

The estimated multispectral endmember signatures $\widehat{\bB}$, as well as the corresponding abundance maps $\widehat{\bS}$, are collectively displayed in Figure \ref{fig: Vancouver Island} (for Vancouver Island data), Figure \ref{fig: Montrose} (for Montrose data) and Figure \ref{fig: Nassau} (for Nassau data).
Using the PRIME algorithm and the Vancouver Island data for example, the estimated $N:=6$ multispectral endmember signatures in $\widehat{\bB}$ (i.e., the $N$ column vectors in $\widehat{\bB}$) are displayed as the $N$ red curves in Figure \ref{fig: Vancouver Island}, and
the estimated $N$ abundance maps in $\widehat{\bS}$ (corresponding to the $N$ rows of $\widehat{\bS}$) are displayed as the $N$ maps in the second row (below GT) in Figure \ref{fig: Vancouver Island}.
This displaying approach also applies to other algorithms and other data (cf. Figure \ref{fig: Montrose} and Figure \ref{fig: Nassau}), and has been frequently used in displaying the BSS results in the HU literature \cite{VCA,HyperCSI}.

For VCA, some of the estimated signatures quite deviate from the GT; see, e.g., the 5th signature in Vancouver Island data (cf. Figure \ref{fig: Vancouver Island}(e)), the 3rd and 6th signatures in Montrose data (cf. Figures \ref{fig: Montrose}(c) and \ref{fig: Montrose}(f)), and the 5th and 6th signatures in Nassau data (cf. Figures \ref{fig: Nassau}(e) and \ref{fig: Nassau}(f)). 
However, VCA can still well estimate some of the signatures, such as the 2nd and 6th signatures in Vancouver Island data (cf. Figures \ref{fig: Vancouver Island}(b) and \ref{fig: Vancouver Island}(f)), the 4th and 5th signatures in Montrose data (cf. Figures \ref{fig: Montrose}(d) and \ref{fig: Montrose}(e)), and the 1st and 4th signatures in Nassau data (cf. Figures \ref{fig: Nassau}(a) and \ref{fig: Nassau}(d)). 
This discovers that part of the multispectral signatures could still be estimated by the vertices of the data convex hull (i.e., simplex in HU).
As for the exact mechanism behind this discovery, it requires more investigation, but it is clear that merely using the convex geometry is insufficient for the underdetermined scenario.

For NMF, some multispectral signatures cannot be even reasonably estimated; see, e.g., the 1st, 3rd and 4th signatures in Vancouver Island data (cf. Figures \ref{fig: Vancouver Island}(a), \ref{fig: Vancouver Island}(c) and \ref{fig: Vancouver Island}(d)), the 1st, 2nd, 3rd and 4th signatures in Montrose data (cf. Figures \ref{fig: Montrose}(a), \ref{fig: Montrose}(b), \ref{fig: Montrose}(c) and \ref{fig: Montrose}(d)), and the 3rd and 6th signatures in Nassau data (cf. Figures \ref{fig: Nassau}(c) and \ref{fig: Nassau}(f)).
Although some signatures exhibit similar shapes to GT (cf. Figures \ref{fig: Montrose}(d)-(f)), their radiance values are quite deviated, showing a fundamental limitation of NMF.
Additionally, NMF shows failure performance when estimating the abundance maps for all the three data, and VCA fails in the the Montrose and Nassau data. 
This is because both methods do not involve any fundamental mechanism to address the underdetermined issue.

With the judicious prism mechanism (cf. Figure \ref{fig:ZhZmrelation}), the induced MU data-fitting term \eqref{eq:DF} can well tackle the highly challenging underdetermined MU problem.
Specifically, for all the three data, it can be seen that the proposed PRIME algorithm achieves highly promising results on all the signature/abundance estimations.
Though the 3rd estimated signature in the Montrose data has some deviation, its curve shape still holds a high resemblance to the GT signature; see the red and black curves in Figure \ref{fig: Montrose}(c).
Accordingly, the 3rd abundance map estimated by PRIME still has very high quality, as can be seen from the first map (GT) and the second map (PRIME) in Figure \ref{fig: Montrose}(c).
The highly remarkable thing is that the proposed PRIME  algorithm successfully separates $6$ sources merely from $4$ observations (i.e., $N>P$), demonstrating the feasibility of MU  for the underdetermined case, under a fully \textit{blind} setting!

To fairly argue the superior performance of PRIME, we also conduct a quantitative evaluation.
The estimated results over three data in terms of SAM, RMSE, and computational time are summarized in Table \ref{tab: quan}, where the boldfaced numbers indicate the smallest SAM/RMSE or the fastest computational time.
For all three data, one can observe that the proposed PRIME algorithm achieves the smallest SAM, indicating the best spectral shape preservation capability in $\widehat{\bB}$, as well as the smallest RMSE, revealing the excellent spatial quality of the estimated abundances in $\widehat{\bS}$.
The computational time of PRIME, however, is longer than other fast algorithms like VCA and NMF, mainly owing to the iterative network update stage (cf. \eqref{update: f}).
Nevertheless, if one accepts a non-blind mechanism, the quantum prism network $f$ may be pre-trained, and we expect that the overall computational time of PRIME would be reduced to just a few seconds.

\begin{table}[t]
    \renewcommand\arraystretch{1.1}
    \caption{Performance comparison among different methods in terms of SAM, RMSE, and computational time.}
    \centering
    
        \begin{tabular}{ c|c||c|c|c }
            \hline
            \hline
            Data &Methods & SAM$~\!(\downarrow)$ & RMSE$~\!(\downarrow)$ & Time (sec.)
            \\
            \hline
            \hline
            \multirow{3}{*}{Vancouver Island} & VCA &13.3257&0.3027  &{\bf 0.0137}
            \\
            &NMF & 12.8905  & 0.2455 & 18.9011 
            \\
            &PRIME & {\bf 6.5405} & {\bf 0.1612} & 54.6156
            \\
            \hline
            \hline
            \multirow{3}{*}{Montrose} & VCA &3.8597 &0.1628 &{\bf 0.0136}
            \\
            &NMF &4.1394  &0.2001 & 16.5596
            \\
            &PRIME & {\bf 2.2104} & {\bf 0.0916} & 54.6125
            \\
            \hline
            \hline
            \multirow{3}{*}{Nassau} & VCA &13.1555 &0.2940 &{\bf 0.0127}
            \\
            &NMF & 20.1224  &0.4466 & 18.4992
            \\
            &PRIME & {\bf 8.1568} & {\bf  0.0830} & 54.9062
            \\
            \hline
            \hline

        \end{tabular}
        
        \label{tab: quan}
\end{table}

\subsection{Ablation Study}\label{sec: ablation}

In this experiment, we conduct an ablation study to investigate the influence of three key components of PRIME.
First, PRIME has adopted the heuristic initialization (HI) strategy \eqref{eq: add noise}.
We will show that this is better than the random initialization strategy when computing $\bZ_h^0$.
Second, we evaluate the value of the SS module in the QUEEN-based prism $f$, which judiciously utilizes the physical meaning of light splitting to reduce half of the network parameters as described in Section \ref{sec:fDesign}.
Otherwise, when SS module is removed, the last layer of the inverse-QC module (i.e., TConvModule 3 defined in Appendix \ref{apx: network}) should be revised to directly have $M=8$ output channels.
Though this seems more straightforward, the performance significantly degrades, as will be seen.
Third, to implement the tough subproblem \eqref{update: AS}, we update $(\bA^t,\bS^t)$ by using the convex geometry (CG) theory \cite{HyperCSI}.
Otherwise, we need to implement \eqref{update: AS} directly using the algebra-based NMF method \cite{NMF-peer}.
Nevertheless, as will be seen, this is less efficient.

We test the proposed PRIME algorithm by turning on two of the three components (and turning off the other one) for all the three data introduced in Section \ref{sec: exp_design}.
The ablation study results averaged over the three data are quantitatively displayed in Table \ref{tab: ablation study}.
We observe that the MU performance of PRIME significantly degrades if SS module or CG module is turned down.
Though the performance of PRIME also degrades when HI module is not used, it is not too significant.
This shows that initialization does play some role when implementing the non-convex MU criterion \eqref{eq:MUcriterion}, but even when $\bZ_h^0$ is simply initialized by uniformly random matrix, the proposed PRIME algorithm still has the capability of retrieving relatively good estimates of $(\widehat{\bB},\widehat{\bS})$ through the subsequent iterative updating procedure (cf. Algorithm \ref{alg: MUmain}).
The proposed PRIME algorithm achieves the best MU performance when all the HI/SS/CG modules are used, proving the effectiveness of these components, as can be seen in Table \ref{tab: ablation study}.

\begin{table}[t]
    \caption{Ablation study for the proposed PRIME algorithm, where the marker ``\CheckmarkBold" means that the corresponding module is used.}
    \vspace{-0.15cm}
    \begin{center}
    \setlength{\tabcolsep}{0.9mm}
    \begin{tabular}{c c c|| c |c}
    \hline
    \hline
    \makecell[c]{HI module}  &\makecell[c]{SS module} &\makecell[c]{CG module} &~SAM$~\!(\downarrow)$ &~RMSE$~\!(\downarrow)$
    \\ 
    % \hline
    %  &\CheckmarkBold &   &24.5926 &2.4855
    % \\
    \hline
    \hline
    \CheckmarkBold  &  \CheckmarkBold & &20.4113 &0.5307
    \\
    % \hline
    %   &    & \CheckmarkBold & 17.7735 & 1.1260
    % \\
    \hline
    \CheckmarkBold  &   &\CheckmarkBold &14.4840 &1.5109
    \\
    \hline
      &\CheckmarkBold  &\CheckmarkBold & 12.1527 &0.1937
    \\
    \hline
    \CheckmarkBold  &\CheckmarkBold  &\CheckmarkBold &{\bf5.6359} &{\bf0.1119}
    \\ 
    \hline
    \hline
    \end{tabular}
    \label{tab: ablation study}
    \end{center}
\end{table}

\begin{figure}[t]
    \begin{center}
    \resizebox{0.45\linewidth}{!}{\hspace{-0.1cm}\includegraphics{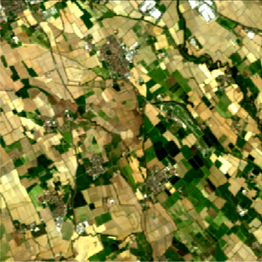}}
    \end{center}
    \vspace{-0.3cm}
    \caption{Region of interest (ROI) of the studied Pavia data acquired by the PRISMA satellite operated by the Agenzia Spaziale Italiana (ASI).}
    \label{fig: PRISMA_GT}
    %\vspace{-0.4cm}
\end{figure}

\begin{figure*}[t]
    \begin{center}
    \resizebox{0.95\linewidth}{!}{\hspace{-0.1cm}\includegraphics{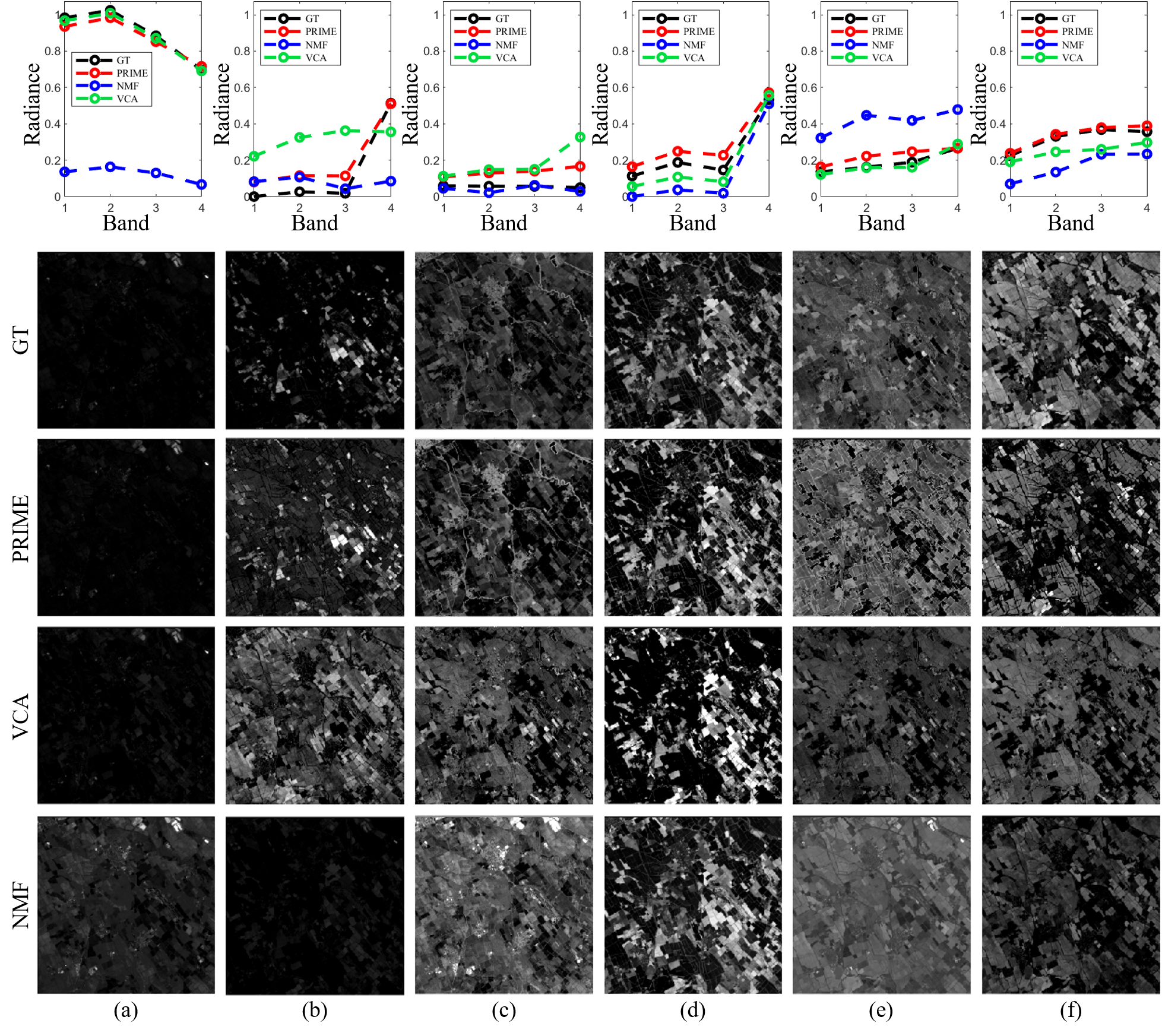}}
    \end{center}
    \vspace{-0.3cm}
    \caption{The MU results, including the estimated abundance maps and the multispectral endmember signatures, for the Pavia data (PRISMA satellite).}
    \label{fig: Pavia}
    \vspace{-0.4cm}
\end{figure*}

\subsection{Related Discussions}\label{sec: discussion}
To further evaluate the performance of PRIME, we use the reference HSI captured by PRecursore IperSpettrale della Missione Applicativa (PRISMA) \cite{PRISMA} satellite, which was launched by the Agenzia Spaziale Italiana (ASI) in 2019.
PRISMA is equipped with $66$ and $170$ contiguous spectral bands in the VNIR and SWIR spectral range, respectively.
As well as Section \ref{sec: exp_design}, the spectral ranges corresponding to the Landsat TM bands 1–4 are used to uniformly downsample the reference signatures, resulting in the $4$-band GT multispectral signatures.
The data (with a spatial resolution of 30 m) was acquired over Pavia, Italy, and with $L=256\times256$ pixels, and the false-color image (composed of band 33 (R), band 21 (G), and band 12 (B)) is displayed in Figure \ref{fig: PRISMA_GT}.

The estimated multispectral signatures $\widehat{\bB}$ and the corresponding abundance maps $\widehat{\bS}$ are collectively displayed in Figure \ref{fig: Pavia}. 
For VCA, the 2nd signature entirely deviates from the GT (cf. Figures \ref{fig: Pavia}(b)). 
In addition, the 3rd, 5th, and 6th abundance maps appear to be highly similar to each other, lacking distinct spatial differences for distinguishing the materials/substances (cf. Figures \ref{fig: Pavia}(c), \ref{fig: Pavia}(e), and \ref{fig: Pavia}(f)).
For NMF, the 1st, 2nd, and 5th signature estimations (along with their respective abundance maps) are fail, exhibiting notable discrepancies from the GT (cf. Figures \ref{fig: Pavia}(a), \ref{fig: Pavia}(b) and \ref{fig: Pavia}(e)).
In contrast, it is evident that all the estimated signatures of PRIME closely resemble the GT.
Moreover, PRIME demonstrates a strong ability to preserve spatial uniqueness across all abundance maps while holding high resemblance to GT.
Table \ref{tab: quan_Pavia} summarizes the quantitative evaluation results, revealing that PRIME outperforms peer methods by achieving the smallest SAM and RMSE.
The evaluation using PRISMA satellite data further underscores the generalizability of PRIME, demonstrating its robustness across diverse sensing platforms.

\begin{table}[t]
    \renewcommand\arraystretch{1.1}
    \caption{Performance comparison among different methods in terms of SAM, RMSE, and computational time.}
    \centering
        \begin{tabular}{ c|c||c|c|c }
            \hline
            \hline
            Data &Methods & SAM$~\!(\downarrow)$ & RMSE$~\!(\downarrow)$ & Time (sec.)
            \\
            \hline
            \hline
            \multirow{3}{*}{Pavia} & VCA &17.2164 &0.2067  &{\bf 0.0162}
            \\
            &NMF &21.8513 &0.1641 & 14.8462
            \\
            &PRIME & {\bf 7.7992} & {\bf 0.1116} &53.4070 
            \\
            \hline
            \hline
        \end{tabular}
        \label{tab: quan_Pavia}
\end{table}

\begin{figure}[t]
    \begin{center}
    \resizebox{0.95\linewidth}{!}{\hspace{-0.1cm}\includegraphics{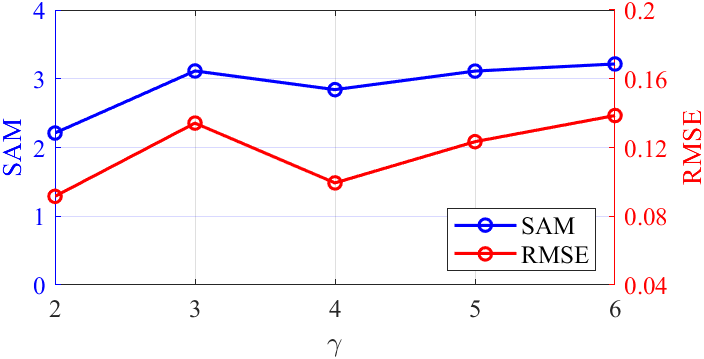}}
    \end{center}
    \vspace{-0.3cm}
    \caption{Performance of PRIME tested across different settings of $\gamma$.}
    \label{fig: gamma}
    %\vspace{-0.4cm}
\end{figure}

Since we mainly focus on unmixing $N=6$ sources from $P=4$ band MSI in Section \ref{sec: algo}, the $\gamma$ is set to $2$, which is sufficient to have $\gamma P > N$.
However, some datasets, such as the Cuprite mining site collected by AVIRIS, may have $N\geq10$ minerals \cite{HyperCSI}, meaning that a larger value of $\gamma$ is needed.
Therefore, we test PRIME on Montrose data to understand the impact of $\gamma$ by largely increasing its value from $2$ to $6$, as demonstrated in Figure \ref{fig: gamma}.
To accommodate the variation, we adjust the methodology accordingly, particularly within the quantum prism network $f$.

As the initial virtual HSI $\bZ_h^0$ is relevant to the $f$ update (cf. Step 4 in Algorithm \ref{alg: MUmain}), we first modify the initialization process (cf. Step 2 in Algorithm \ref{alg: MUmain}) so that each multispectral band in $\bZ_m$ is divided into $\gamma$ hyperspectral bands, while maintaining the light-splitting property with slight perturbation (cf. Appendix \ref{apx: init}).
Next, we revise the last layer of the inverse-QC module (i.e., TConvModule 3, as detailed in Appendix \ref{apx: network}) to output ($\gamma P -P$) bands, while the remaining $P$ bands were calculated within the SS module based on the light-splitting relation.
To maintain the spectral continuity of the virtual HSI, as $\gamma$ increased from $2$ to $6$ (cf. Figure \ref{fig: gamma}), we progressively adjusted $\alpha$ to 1E-4 (for $\gamma=2$), 2E-4 (for $\gamma=3$), 5E-4 (for $\gamma=4$), 6E-4 (for $\gamma=5$) and 1E-3 (for $\gamma=6$).
%
%$\{1\text{E}-4, 2\text{E}-4, 5\text{E}-4, 6\text{E}-4, 1\text{E}-3 \}$
%
We observe that SAM and RMSE only have subtle variations as $\gamma$ increases, meaning that PRIME is able to address a more serious underdetermined BSS scenario that requires higher $\gamma$.
In addition, the computational time approximately ranges from $54$ to $64$ seconds, indicating that the impact of $\gamma$ on the computational efficiency is ignorable.
These findings highlight the robustness of the proposed PRIME algorithm, making it applicable for underdetermined MU scenarios requiring higher value of $\gamma$.

% \begin{table}[h]
%     \renewcommand\arraystretch{1.1}
%     \caption{Performance comparison of different $\gamma$ on Montrose data.}
%     \centering
%         \begin{tabular}{c|c||c|c|c }
%             \hline
%             \hline
%             $\gamma$ &$\alpha$ &SAM$~\!(\downarrow)$ & RMSE$~\!(\downarrow)$ & Time (sec.)
%             \\
%             \hline
%             \hline
%             2 &1\text{E}-4 &2.2104 &0.0916 &54.6125
%             \\
%             3 &2\text{E}-4 &3.1136 &0.1342 &56.5090
%             \\
%             4 &5\text{E}-4 &2.8416 &0.0994 &59.0708
%             \\
%             5 &6\text{E}-4 &3.1115 &0.1234 &61.2201
%             \\
%             6 &1\text{E}-3 &3.2152 &0.1386 &64.6486
%             \\
%             \hline
%             \hline

%         \end{tabular}
%         \label{tab: gamma}
%     \end{table}
%The HSI, captured by ITRES CASI 1500 imager, covers a 380–1050-nm spectral range with 48 bands at a 1-m ground sample distance (GSD).

\section{Conclusion}\label{sec: conclusion}

In this work, we have the first attempt to resolve the MU problem for the underdetermined scenario, where we have more sources than the MSI bands.
The proposed PRIME algorithm judiciously introduces a virtual quantum prism to tackle the underdetermined issue.
The prism conducts the light-splitting task to generate more virtual bands, and induces a novel data-fitting term for MU.
The data-fitting term, together with some customized geometry-based volume and algebra-based sparsity regularization terms, forms the proposed MU criterion.
Our PRIME algorithm elegantly implements the criterion by plugging in the computationally efficient convex geometry method, and proposes a quantum deep image prior (QDIP) strategy to achieve a fully \textit{blind} mechanism.
For the virtual quantum prism, the adopted quantum feature extraction technique has the mathematical guarantee of quantum full expressibility.
To evaluate the MU performance, we also design the first MU experimental protocol that can be easily implemented for future investigators.
As it turns out, the PRIME algorithm is highly effective for addressing the MU problem in terms of both quantitative and qualitative perspectives.

\appendix
\subsection{Initialization of PRIME Algorithm} \label{apx: init}
Since the network $f$ will be updated first in Algorithm \ref{alg: MUmain}, we need to initialize $\bZ_h$ and $(\bA,\bS)$.
For $\bZ_h$, we propose to initialize it as a slightly perturbed light-splitting version of the $P$-band $\bZ_m$, as detailed below.
As mentioned in Section \ref{sec: intro}, the virtual $\bZ_h$ has no need to correspond to some real HSI sensor, so we can flexibly assume that each band in $\bZ_m$ is split into $\gamma\in\mathbb{Z}_{++}$ bands (i.e., $M=\gamma P$, where $\gamma$ is equivalent to upsampling factor in spectral super-resolution task).

Therefore, the spectral response matrix $\bD$ is given by $\bD\triangleq\bI_P\otimes\bm{1}^T_\gamma\in\mathbb{R}^{P\times M}$.
Considering that the concept of the $M$-band virtual $\bZ_h$ is proposed to mitigate the underdetermined issue (thus requiring $M=\gamma P\geq N$), and that we usually have $2P\geq N$ in practice as discussed before, we can simply consider $\gamma:=2$ for conciseness.
According to the physical meaning of light splitting, the $i$th MSI band $[\bZ_m]_{(i,:)}$ can be split into two HSI bands, i.e., $0.5([\bZ_m]_{(i,:)}\pm {[\bm\theta]}_{(i,:)})$ for some $[\bm\theta]_{(i,:)}$, which ensures the light-splitting relation of $\bD\bZ_h=\bZ_m$.
Considering the desired continuity of the HSI bands, we can set $[\bm\theta]_{(i,:)}$ as $\frac{1}{4}([\bZ_m]_{(i+1,:)}-[\bZ_m]_{(i,:)})$ for uniform spectral sampling; further considering the boundary issue, the last term (i.e., the $P$th term) is set as $[\bm\theta]_{(P,:)}$ as $\frac{1}{4}([\bZ_m]_{(P,:)}-[\bZ_m]_{(P-1,:)})$.
To be precise, the above generated $M$-band information $\widetilde{\bZ}_h$ can be explicitly written as
\begin{align*}
\left[\widetilde{\bZ}_h\right]_{(i,:)}=\left\{
\begin{aligned}
&0.5\left( \left[\bZ_m\right]_{(\frac{i+1}{2},:)}-\left[{\bm\theta}\right]_{(\frac{i+1}{2},:)} \right),&\!\!\!\!\textrm{for odd}~i,
\\
&0.5\left( \left[\bZ_m\right]_{(\frac{i}{2},:)}+\left[{\bm\theta}\right]_{(\frac{i}{2},:)} \right),&\!\!\!\!\!\!\!\textrm{for even}~i.
\end{aligned}
\right.
\end{align*}
It can be proven true that the $M$-band $\widetilde{\bZ}_h$ is still of rank $P$, so it is slightly perturbed with Gaussian noise $\bN$ whose energy is about $e:=5\%$ of $\widetilde{\bZ}_h$, leading to the initialization of
\begin{align}\label{eq: add noise}
\bZ_h^0= \widetilde{\bZ}_h+c\bN,~\textrm{with}~c\triangleq \sqrt{\frac{e\|\widetilde{\bZ}_h\|_F^2}{\|\bN\|_F^2}},
\end{align}
followed by projecting it back to the non-negative orthant to meet the non-negative nature of HSI.
Finally, considering that the slightly perturbed version is empirically observed to be ill-conditioned, though of full rank, we initialize $(\bA^0,\bS^0)$ by performing HU on the virtual HSI $\bZ_h^0$ using the John ellipsoid-driven BSS method \cite{HISUN} called highly-mixed/ill-conditioned spectrum unmixing (HISUN), which is proven to be robust against the ill-conditioned issue.
The above heuristic initialization strategy will be experimentally proven effective, and is summarized in Algorithm \ref{alg: MUmain}.

\subsection{Detailed Design of Quantum Prism Network $f$} \label{apx: network}

The implementation details of the quantum prism network $f$ is organized in Table \ref{tab: configuration}.
To precisely describe the network architecture, we define the used configurations.
Some frequently used quantum neurons and their mathematical definitions are summarized in 
Table \ref{tab: common_qu_gate}.
Also, ``CB($o, k, p$)" denotes convolution block composed of the convolution layer ``Conv($o, k, p$)" and LeakyReLU (with a negative slope $0.2$), while ``TCB($o, k$)" denotes transposed CB composed of transposed convolution TConv($o, k$) and LeakyReLU, where ($o, k, p$) specify the number of output channels, the kernel size of $k\times k$, and the number of paddings, respectively.
The ``MaxPool" denotes the Max Pooling, the ``Up-sample" is implemented by bilinear interpolation, and the non-negative projection (``NNProj") is implemented by ReLU, respectively.
The core signal processing in the quantum prism network $f$ is done by quantum computing and learning.
The quantum neurons, also known as quantum gates, include those rotation gates, Ising gates and Toffoli gates \cite{HyperQUEEN}.
The gates $``R_Y$($g, n$)$," ``R_X$($g, n$)$," ``XX$($g, n$)$,"$ and $``Z$($g, n$)$"$ are defined in Table \ref{tab: common_qu_gate}, where $g$ and $n$ respectively denote the number of parallel groups and the number of qubits in each group.
In Toffoli entanglement layer, the $``\text{CCNOT}$($cw_1, cw_2, tw$)$"$ gate is employed to capture the quantum entanglement effect, where $cw_1$, $cw_2$, and $tw$ specify the wires involved: $cw_1$ is the first control wire, $cw_2$ is the second control wire, and $tw$ is the target wire on which the operation acts.

\begin{center}
\scriptsize
\begin{table}[ht]
    \caption{Detailed architecture of the proposed quantum prism $f$.}
    \centering
    \label{tab: configuration}
\begin{tabular}{c||c|c|c}
    \hline
    \hline
    \rule{0pt}{2.3ex}
     & Layer & Configuration & Output Size
    \rule{0pt}{2ex}
     \\
    \hline
    \hline
    \rule{0pt}{2.3ex}
    \multirow{1}{*}{} & Input & - & 4$\times$256$\times$256
    \rule{0pt}{2ex}
    \\
    \hline
    \rule{0pt}{2.3ex}
    \multirow{8}{*}{\makecell{DC Module}} & \multirow{2}{*}{ ConvModule 1}  & CB(8, 3, 1)$\times$1 & \multirow{2}{*}{8$\times$252$\times$252}
    \rule{0pt}{2.3ex}\\
     & & CB(8, 3, 0)$\times$2  & 
    \rule{0pt}{2.3ex}\\
    \cline{2-4} &  \multirow{2}{*}{ConvModule 2} &  2$\times$2 MaxPool & \multirow{2}{*}{8$\times$120$\times$120}
    \rule{0pt}{2.3ex}\\
     &  & CB(8, 3, 0)$\times$3 & 
    \rule{0pt}{2.3ex}\\
    \cline{2-4} &  \multirow{2}{*}{ConvModule 3} &  2$\times$2 MaxPool & \multirow{2}{*}{8$\times$54$\times$54}
    \rule{0pt}{2.3ex}\\
     &  & CB(8, 3, 0)$\times$3 & 
    \rule{0pt}{2.3ex}\\
    \hline
    \multirow{21}{*}{\makecell{Core\\Quantum\\FE Module\\(\textbf{Theorem \ref{theorem: FE}})}} & Reshape  & - & 2(54$^2$)$\times$4
    \rule{0pt}{2.3ex}\\
    \cline{2-4} & \makecell{Angle Embedding} & $R_{Y}$(2(54$^2$), 4) & 2(54$^2$)$\times$4
    \rule{0pt}{2.3ex}\\
    \cline{2-4} & Unitary Gate 1 & $R_{Y}$(2(54$^2$), 4) & 2(54$^2$)$\times$4
    \rule{0pt}{2.3ex}\\
    \cline{2-4} & Reshape  & - & 4(54$^2$)$\times$2
    \rule{0pt}{2.3ex}\\
    \cline{2-4} & Unitary Gate 2 & $XX$(4(54$^2$), 2) & 4(54$^2$)$\times$2
    \rule{0pt}{2.3ex}\\
    \cline{2-4} & Reshape  & - & 2(54$^2$)$\times$4
    \rule{0pt}{2.3ex}\\
    \cline{2-4} & Unitary Gate 3 & $R_{X}$(2(54$^2$), 4) & 2(54$^2$)$\times$4
    \rule{0pt}{2.3ex}\\
    \cline{2-4} & Reshape  & - & 4(54$^2$)$\times$2
    \rule{0pt}{2.3ex}\\
    \cline{2-4} & Unitary Gate 4 & $XX$(4(54$^2$), 2) & 4(54$^2$)$\times$2
    \rule{0pt}{2.3ex}\\
    \cline{2-4} & Reshape  & - & 2(54$^2$)$\times$4
    \rule{0pt}{2.3ex}\\
    \cline{2-4} & Unitary Gate 5 & $R_{Y}$(2(54$^2$), 4) & 2(54$^2$)$\times$4
    \rule{0pt}{2.3ex}\\
    \cline{2-4} & \multirow{4}{*}{\makecell{Toffoli\\Entanglement}} & CCNOT(0, 1, 2) & 2(54$^2$)$\times$4
    \rule{0pt}{2.3ex}\\
    \cline{3-4} &  & CCNOT(1, 2, 3) & 2(54$^2$)$\times$4
    \rule{0pt}{2.3ex}\\
    \cline{3-4} &  & CCNOT(2, 3, 0) & 2(54$^2$)$\times$4
    \rule{0pt}{2.3ex}\\
    \cline{3-4} &  & CCNOT(3, 0, 1) & 2(54$^2$)$\times$4
    \rule{0pt}{2.3ex}\\
    \cline{2-4} & \makecell{QC\\Measurement} & $Z$(2(54$^2$), 2) & 2(54$^2$)$\times$2
    \rule{0pt}{3ex}\\
    \cline{2-4} & Reshape  & - & 4$\times$54$\times$54
    \rule{0pt}{2.3ex}\\
    
    \hline
    \multirow{7}{*}{\makecell{Inverse-QC\\Module}} & \multirow{2}{*}{TConvModule 1} & TCB(8, 3)$\times$3 & \multirow{2}{*}{8$\times$120$\times$120}
    \rule{0pt}{2.3ex}\\
    & & 2$\times$2 Up-sample &
    \rule{0pt}{2.3ex}\\
    \cline{2-4} & \multirow{2}{*}{TConvModule 2} & TCB(8, 3)$\times$3 & \multirow{2}{*}{8$\times$252$\times$252}
    \rule{0pt}{2.3ex}\\
    & & 2$\times$2 Up-sample &
    \rule{0pt}{2.3ex}\\
    \cline{2-4} & \multirow{2}{*}{TConvModule 3} & TCB(8, 3)$\times$1 & \multirow{2}{*}{4$\times$256$\times$256}
    \rule{0pt}{2.3ex}\\
     &  & TCB(4, 3)$\times$1 & 
    \rule{0pt}{2.3ex}\\
    
    \hline
    \multirow{3}{*}{\makecell{SS Module}} & \multirow{1}{*}{Subtraction} & - & \multirow{1}{*}{4$\times$256$\times$256}
    \rule{0pt}{2.3ex}\\
    \cline{2-4} & \multirow{1}{*}{Insertion} & Alternating & \multirow{1}{*}{8$\times$256$\times$256}
    \rule{0pt}{2.3ex}\\
    \cline{2-4} & \multirow{1}{*}{Projection} &  NNProj & \multirow{1}{*}{8$\times$256$\times$256}
    \rule{0pt}{2.3ex}\\
    \hline
\end{tabular}
\end{table}
%\vspace{-0.5cm}
\end{center}

\renewcommand{\thesubsection}{\Alph{subsection}}
\bibliography{ref}

\begin{IEEEbiography}[{\resizebox{0.9in}{!}{\includegraphics[width=1in,height=1.25in,clip,keepaspectratio]{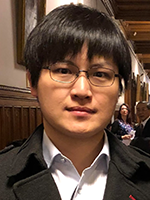}}}]
{\bf Chia-Hsiang Lin}
(S'10-M'18-SM'24)
received the B.S. degree in electrical engineering and the Ph.D. degree in communications engineering from National Tsing Hua University (NTHU), Taiwan, in 2010 and 2016, respectively.
From 2015 to 2016, he was a Visiting Student of Virginia Tech,
Arlington, VA, USA.

He is currently an Associate Professor with the Department of Electrical Engineering, and also with
the Miin Wu School of Computing,
National Cheng Kung University (NCKU), Taiwan.
Before joining NCKU, he held research positions with The Chinese University of Hong Kong, HK (2014 and 2017),
NTHU (2016-2017),
and the University of Lisbon (ULisboa), Lisbon, Portugal (2017-2018).
He was an Assistant Professor with the Center for Space and Remote Sensing Research, National Central University, Taiwan, in 2018, and a Visiting Professor with ULisboa, in 2019.
His research interests include network science,
quantum computing,
convex geometry and optimization, blind signal processing, and imaging science.

Dr. Lin received the Emerging Young Scholar Award (The 2030 Cross-Generation Program) from National Science and Technology Council (NSTC), from 2023 to 2027,
the Future Technology Award from NSTC, in 2022,
the Outstanding Youth Electrical Engineer Award from The Chinese Institute of Electrical Engineering (CIEE), in 2022,
the Best Young Professional Member Award from IEEE Tainan Section, in 2021,
the Prize Paper Award from IEEE Geoscience and Remote Sensing Society (GRS-S), in 2020,
the Top Performance Award from Social Media Prediction Challenge at ACM Multimedia, in 2020,
and The 3rd Place from AIM Real World Super-Resolution Challenge at IEEE International Conference on Computer Vision (ICCV), in 2019.
He received the Ministry of Science and Technology (MOST) Young Scholar Fellowship, together with the EINSTEIN Grant Award, from 2018 to 2023.
In 2016, he was a recipient of the Outstanding Doctoral Dissertation Award from the Chinese Image Processing and Pattern Recognition Society and the Best Doctoral Dissertation Award from the IEEE GRS-S.
\end{IEEEbiography}

\begin{IEEEbiography}[{\resizebox{1in}{!}{\includegraphics[width=1in,height=1.25in,clip,keepaspectratio]{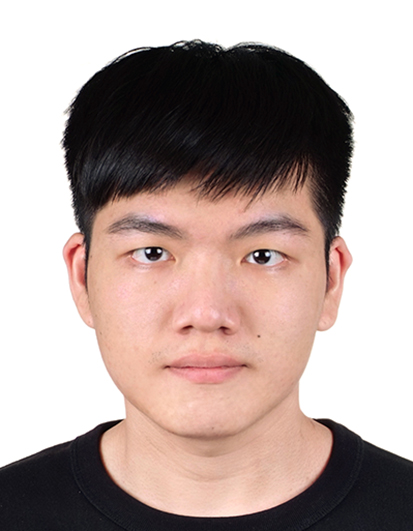}}}]
    	{\bf Jhao-Ting Lin}
		(S'20)
received his B.S. degree from the Department of Communications, Navigation and Control Engineering, National Taiwan Ocean University, Taiwan, in 2020.

He is currently a Ph.D. student affiliated with the Intelligent Hyperspectral Computing Laboratory, Department of Electrical Engineering, National Cheng Kung University, Taiwan. 
His research interests include convex optimization, signal processing, quantum computing, and hyperspectral imaging.

He has received some highly competitive student awards, including the 2022 and 2024 Pan Wen Yuan Award from the Industrial Technology Research Institute (ITRI) of Taiwan.
He has been selected as a recipient for the Ph.D. Students Study Abroad Program from the National Science and Technology Council (NSTC), Taiwan, in 2025.
\end{IEEEbiography}

\end{document}